\documentclass[letterpaper]{article} 
\usepackage{aaai25}  
\usepackage{times}  
\usepackage{helvet}  
\usepackage{courier}  
\usepackage[hyphens]{url}  
\usepackage{graphicx} 
\usepackage{natbib}  
\usepackage{caption} 
\usepackage{algorithm}
\usepackage{algorithmic}

\usepackage{amsmath,amsfonts,amsthm}
\newtheorem{theorem}{Theorem}

\usepackage{subfig}                 
\usepackage{overpic}                
\usepackage{multirow}               
\usepackage{booktabs}               
\usepackage{hhline}                 

\usepackage{newfloat}
\usepackage{listings}
\DeclareCaptionStyle{ruled}{labelfont=normalfont,labelsep=colon,strut=off} 
\floatstyle{ruled}
\newfloat{listing}{tb}{lst}{}
\floatname{listing}{Listing}
\title{Entire-Space Variational Information Exploitation for\\ Post-Click Conversion Rate Prediction}
\author{
    Ke Fei,
    Xinyue Zhang,
    Jingjing Li\thanks{Corresponding author}
}
\affiliations{
    University of Electronic Science and Technology of China.\\
    Chengdu, China\\
    q191106702@gmail.com, qzmvc1@gmail.com, lijin117@yeah.net
}

\usepackage{bibentry}

\begin{document}

\maketitle

\begin{abstract}
In recommender systems, post-click conversion rate (CVR) estimation is an essential task to model user preferences for items and estimate the value of recommendations. Sample selection bias (SSB) and data sparsity (DS) are two persistent challenges for post-click conversion rate (CVR) estimation. Currently, entire-space approaches that exploit unclicked samples through knowledge distillation are promising to mitigate SSB and DS simultaneously. Existing methods use non-conversion, conversion, or adaptive conversion predictors to generate pseudo labels for unclicked samples. However, they fail to consider the unbiasedness and information limitations of these pseudo labels. Motivated by such analysis, we propose an entire-space variational information exploitation framework (EVI) for CVR prediction. First, EVI uses a conditional entire-space CVR teacher to generate unbiased pseudo labels. Then, it applies variational information exploitation and logit distillation to transfer non-click space information to the target CVR estimator. We conduct extensive offline experiments on six large-scale datasets. EVI demonstrated a 2.25\% average improvement compared to the state-of-the-art baselines.
\end{abstract}

%
\begin{links}
    \link{Code}{https://github.com/q1179897215/EVI}
\end{links}

\section{INTRODUCTION}
Modern recommender systems aim to connect users with the items they want using efficient computation. Estimating the post-click conversion rate (CVR) is a fundamental task that models the interaction between users and items, such as user preferences for specific items. \cite{xi_modeling_2021, ma_entire_2018, dai2022generalized}. The na\"ive CVR model is trained using clicked samples from the click space, but it needs to estimate CVR for all impression items from the entire impression space, which includes unclicked samples. Since click and conversion events are interconnected, unclicked samples are not missing at random (MNAR) \cite{marlin2009collaborative, zhang_large-scale_2020}. This gap between the click space and the entire space leads to sample selection bias (SSB) \cite{ma_entire_2018, li2023balancing}. Moreover, the data sparsity (DS) of clicked samples and converted samples further hinder the generalization of click-space CVR models to entire space \cite{ wang_escm2_2022}.

To mitigate the SSB, causal inference techniques such as inverse propensity scores (IPS) and double robust learning (DR) have been proposed to debias the CVR estimation and control variance \cite{zhang_large-scale_2020, guo_enhanced_2021, dai2022generalized}.
The essence of IPS and DR is to reconstruct the entire space loss by weighting the clicked samples' losses with IPS.
    
Since causal methods are trained only on click space, exploring the entire space is a promising approach to mitigate SSB and DS simultaneously \cite{su2024ddpo}. Initially, to utilize unclicked samples, Ma et al. \cite{ma_entire_2018} proposed ESMM to model CVR on the entire space through learning a click-through and conversion rate (CTCVR) task. Based on the chain rule of probability, CTCVR = click-through rate (CTR) * CVR. Wang et al. \cite{wang_escm2_2022} combine IPS and DR with ESMM to debias the CVR estimator and simultaneously exploit the entire space information. Xu et al. \cite{xu2022ukd} proposed UKD, which trains an adaptive CVR teacher through adversarial learning to generate soft labels for unclicked samples. These pseudo labels allow the CVR estimator to be trained on the entire space using a distillation approach. In DCMT \cite{zhu2023dcmt}, Zhu et al. labels unclicked samples as non-conversions and apply inverse non-click propensity to correct the bias in the non-conversion estimators. Similar to UKD and DCMT, Su et al. \cite{su2024ddpo} train a na\"ive CVR estimator to produce pseudo labels for unclicked samples.

From a causal inference perspective \cite{rubin2005causal}, we must measure all variables that influence clicks and conversions (also called confounders) to ensure unbiased CVR estimat. However, most existing entire-space methods ignore click information (which contains confounders) when generating pseudo labels, thus transferring the bias to the target estimator. UKD employs adversarial learning to mitigate the domain shift between the click and non-click spaces, but it also results in the loss of click information. DDPO uses the na\"ive CVR estimator to generate pseudo CVR labels for unclicked samples, neglecting click information when training the CVR estimator.

Since the information carried by non-click space pseudo soft labels is easily biased and limited, it is beneficial to construct an unbiased CVR teacher and transfer information at both the logit and feature levels. In this paper, we propose an entire-space variational information exploitation
framework (EVI) for CVR prediction. Firstly, we use click propensities to condition the original sample representations. Based on the click-conditioned representations, we then train an unbiased CVR teacher. To better utilize non-click spaces, we maximize the variational information between the CVR teacher and the CVR student, and we perform logit distillation using unbiased pseudo labels. Our contributions are highlighted as follows:
\begin{itemize}
    \item We introduce EVI, a novel entire-space approach for CVR prediction that avoids bias from pseudo labels and enhances the exploitation of non-click space using features and logits distillation.
    \item We validate the effectiveness of our proposed EVI on five public and one industrial dataset. In general, EVI shows an average AUC improvement of 2.25\% and a 2.78\% NLL reduction over the optimal baselines.
    \item We demonstrate both theoretically and experimentally that the unbiasedness of EVI can be achieved. Additionally, we show that conditional entire-space CVR teacher and variational information exploitation can significantly reduce the mean bias of the CVR estimate.

\end{itemize}

\section{RELATED WORK}
\subsection{Multi-task Learning}
In recommender systems, click events directly affect conversion events, indicating the existence of common variables (confounders) that affect both. CTR and CVR tasks are jointly optimized in a unified multi-task learning (MTL) framework for exploit these confounders.

Based on the intuition from the Mixture of Experts (MoE) approach \cite{jacobs1991adaptive}, Ma et al. \cite{ma_modeling_2018} proposed MMOE, which uses multiple expert networks and task-specific gates to combine task features for CTR and CVR towers. The aim of MMOE is to reduce negative transfer between tasks. While MMOE attempts to solve the issue of negative transfer, Xi et al. proposed AITM \cite{xi_modeling_2021}, which uses an attention mechanism to transfer information from clicks to conversions. Several studies \cite{ma_entire_2018, zhang2022ctnocvr, wen2021hierarchically, wen2020entire} introduce special auxiliary tasks to transfer new information to the target tasks.

\subsection{Causal Inference}

Common multi-task frameworks exploit confounders' information through parameter sharing or attention layers. Causal methods employ pCTR as a compression of confounders' information to handle the SSB problem in CVR estimation \cite{luo2024survey}. Causal inference techniques, such as IPS and DR \cite{pearl2009causal} are employed to the remove the SSB \cite{zhang_large-scale_2020, saito2020doubly, guo_enhanced_2021, saito2020unbiased, wang2022unbiased, wang_escm2_2022, schnabel2016recommendations} in CVR task. Causal methods weight the CVR loss using predicted inverse propensity scores, which are easily affected by high variance issues. SNIP \cite{schnabel2016recommendations}, Clipped-IPS \cite{saito2020unbiased}, MRDR \cite{guo_enhanced_2021}, DR-MSE \cite{dai2022generalized}, DR-BIAS \cite{dai2022generalized}, MRDR-GPL \cite{zhou2023generalized}, and DR-V2\cite{li2023propensity} have been proposed to control the variance.

\section{PRELIMINARY}
\subsection{Problem Formulation}

In online recommender systems, we collect data on user attributes, item attributes, and their interactions (such as clicks and conversions) from the entire impression space, also referred to as the entire space. This data allows us to train CTR and CVR estimators simultaneously using a multi-task framework. Let $\mathcal{U} = \{u_1, u_2,\ldots,u_n\}$ be the set of users and $\mathcal{I} = \{i_1, i_2,\ldots,i_n\}$ be the set of items. The entire space is delineated by the set $\mathcal{D} = \mathcal{U} \times \mathcal{I}$, where $\times$ denotes the Cartesian product operation. Let $o_{u,i}$ be the true click label for the user-item pair $(u, i)$, which can be either 0 or 1. Therefore, the entire space $\mathcal{D}$ is divided into two subsets: the click space $\mathcal{O}$, where $o_{u,i}=1$, and the non-click space $\mathcal{N}$, where $o_{u,i}=0$. The predicted click-through rate (pCTR) or click propensity, $\hat{o}_{u,i} \in [0,1]$, is estimated by a model $f(\cdot;\theta_{\mathrm{CTR}})$ trained on $\mathcal{D}$ with parameters $\theta_{\mathrm{CTR}}$. Let $r_{u,i} \in \{0,1\}$ denote the true post-click labels. The predicted post-click conversion rate (pCVR), $\hat{r}_{u,i} \in [0,1]$, is estimated by models $f(\cdot; \theta_{\mathrm{CVR}})$ with parameters $\theta_{\mathrm{CVR}}$. Let $\mathbf{x}_{u,i}$ represent the feature embeddings of the user-item pair $(u, i) \in \mathcal{D}$.

\subsection{Entire-Space CVR Estimation}

Given that the CVR estimator is required to predict the pCVR in the entire space $\mathcal{D}$, the ideal loss function for the CVR estimator is:

\begin{align}
\mathcal{L}^{ideal}_{\mathrm{CVR}} = \frac{1}{|\mathcal{D}|} \sum_{(u,i) \in \mathcal{D}} \delta(r_{u,i}, \hat{r}_{u,i}), 
\end{align}
where $\delta(\cdot,\cdot)$ denotes a specific loss function, commonly cross-entropy loss function \cite{mao2023cross}. Since the CVR labels of unclicked samples are missing, the na\"ive CVR estimator can only be trained using clicked samples. The training loss for the na\"ive CVR estimator is defined as:

\begin{equation}
\mathcal{L}_{\mathrm{CVR}}^{naive} = \frac{1}{|\mathcal{D}|} \sum_{(u,i) \in \mathcal{D}} o_{u,i}\delta(r_{u,i}, \hat{r}_{u,i}).
\end{equation}

However, these CVR labels of unclicked samples are missing not at random (MNAR) \cite{marlin2009collaborative, zhang_large-scale_2020}, leading to SSB, i.e.,  
\begin{equation} |\mathbb{E}_{\mathcal{D}}\left\lbrack\mathcal{L}^{naive}_{\mathrm{CVR}}\right\rbrack-\mathcal{L}^{ideal}_{\mathrm{CVR}}|>0.
\end{equation}

SSB originates from the interdependence between clicks and conversions. This means that post-click conversion events and click events are both influenced by some common variables (confounders). From a causal inference perspective \cite{rubin2005causal}, we must measure all confounders to ensure the unbiasedness of the CVR estimator. Co-training the CTR estimator helps the CVR estimator capture confounding information, improving the debiasing effect. The effectiveness of multi-task learning has been verified in previous works such as MMOE \cite{ma_modeling_2018} and AITM \cite{xi_modeling_2021}. The CTR task loss is formulated as:
\begin{align}
\mathcal{L}_{\mathrm{CTR}} = \frac{1}{|\mathcal{D}|} \sum_{(u,i) \in \mathcal{D}} \delta(o_{u,i}, \hat{o}_{u,i}).
\end{align}

The ESMM \cite{ma_entire_2018} addresses SSB by introducing an entire-space CTCVR task, which directly connects CVR to CTR based on the chain rule of probability \cite{feller1991introduction}. The ESMM extends the na\"ive CVR estimator to the entire-space CVR estimator, and the CTCVR loss is formulated as:

\begin{equation}
\mathcal{L}_{\mathrm{CTCVR}} = \frac{1}{|\mathcal{D}|} \sum_{(u,i) \in \mathcal{D}} \delta(o_{u,i}*r_{u,i}, \hat{o}_{u,i}*\hat{r}_{u,i}).
\end{equation}

ESCM$^2$ integrates IPW and DR with ESMM. The ESCM$^2$-IPW and ESCM$^2$-DR estimators are theoretically unbiased if the CTR estimation is accurate. The main difference between ESCM$^2$ and ESMM lies in the CVR loss:

\begin{align}
\mathcal{L}_{\mathrm{CVR}}^{\mathrm{IPW}}=\frac1{|\mathcal{D}|}\sum_{(u,i)\in\mathcal{D}}\frac{o_{u,i}\delta_{}(r_{u,i},\hat{r}_{u,i})}{\hat{o}_{u,i}},
\end{align}

\begin{align}
\mathcal{L}_{\mathrm{CVR}}^{\mathrm{DR}}=\frac1{|\mathcal{D}|}\sum_{(u,i)\in\mathcal{D}}\Big[\hat{e}_{u,i}+\frac{o_{u,i}(e_{u,i}-\hat{e}_{u,i})}{\hat{o}_{u,i}}\Big],
\end{align}
where $e_{u,i}=\delta(r_{u,i}, \hat{r}_{u,i})$ and $\hat{e}_{u,i}$ is predicted $e_{u,i}$ by an imputation model with parameter $f(\cdot;\theta_{\mathrm{IMP}})$. 

Unlike ESMM, which uses CTCVR estimation, UKD \cite{xu2022ukd} employs a click-adaptive CVR teacher to produce pseudo conversion labels $r_{u,i}^{a}$ for unclicked samples. UKD enables training the CVR estimator on the entire space. The CVR loss of UKD is formulated as:
\begin{align}
&\mathcal{L}^{\mathrm{UKD}}_{\mathrm{CVR}}=\notag \\ &\frac{1}{|\mathcal{D}|}\sum_{(u,i)\in\mathcal{D}}\left\lbrack o_{u,i}\delta(r_{u,i},\hat{r}_{u,i})\right\rbrack +\left(1-o_{u,i})\delta(r_{u,i}^{a},\hat{r}_{u,i}^{}\right)],\end{align}

UKD trains a click-adaptive CVR teacher using adversarial domain adaptation techniques like gradient reversal \cite{ganin2016domain}. This approach enables the CVR teacher to generate click-adaptive representations that can deceive a click discriminator, rendering it unable to distinguish between clicked and unclicked samples. However, 'click-adaptive' also implies the loss of click information, which includes all confounding information. 
The loss of confounding information leads to SSB in the CVR teacher, which in turn gets transferred to the CVR student (see Theorem 1).

Similar to UKD, DDPO \cite{su2024ddpo} constructs a na\"ive CVR teacher trained on click space to generate pseudo conversion labels $r_{u,i}^{c}$ for unclicked samples. Additionally, DDPO uses inverse click propensity scores and inverse non-click propensity scores to weight the losses of clicked and unclicked samples, respectively.
\begin{equation}
\begin{aligned} 
& \mathcal{L}_{\mathrm{CVR}}^{\mathrm{DDPO}}= \notag \\ & \frac{1}{|\mathcal{D}|}\sum_{(u,i)\in\mathcal{D}} \left \lbrack \frac{o_{u,i}\delta(r_{u,i},\hat{r}_{u,i})}{\hat{o}_{u,i}}+\frac{\left(1-o_{u,i})\delta(r_{u,i}^{c},\hat{r}_{u,i}\right)}{1-\hat{o}_{u,i}}\right\rbrack
\end{aligned}
\end{equation}

DDPO is based on a strong assumption (see Theorem 1 in DDPO original paper) that the CVR estimate of DDPO is unbiased when pseudo labels produced CVR teacher are accurate. However, the CVR teacher of DDPO are trained on click space, which is prone to SSB.

\begin{theorem}
    The entire-space CVR estimator is biased when the pseudo conversion labels of unclicked samples $r^*_{u,i}$ are biased, i.e., \begin{equation}|\mathbb{E}_{\mathcal{D}}\left\lbrack\mathcal{L}^{entire}_{CVR}\right\rbrack-\mathcal{L}^{ideal}_{CVR}|>0. \notag
    \end{equation}
\end{theorem}

\begin{proof}
See Appendix B.1.
\end{proof}

In general, the main aim of entire-space CVR modeling is to leverage the non-click space information to combat SSB and DS. However, pseudo labels carry limited information and are easily biased due to the loss of click information. In this paper, we aim to generate unbiased pseudo labels and enhance non-click space exploitation through variational information exploitation and logit distillation. 

\section{PROPOSED METHOD}
\subsection{EVI Framework}

As discussed in Section 3, the pseudo labels carry limited information and are easily biased due to the loss of click information in CVR teacher. Motivated by this, we propose EVI. As shown in Figure \ref{Framework}, we construct an entire-space CVR teacher trained on click-conditioned representations to ensure the unbiasedness of the CVR teacher. To enhance the entire-space knowledge transfer from the CVR teacher to the target CVR estimator (student), we perform two main actions: 1) use the conditional entire-space CVR teacher to produce unbiased pseudo conversion labels for logit distillation,  and 2) maximize the variational information between the CVR teacher and the CVR student.

\subsection{Conditional Entire-Space CVR Teacher}

In Section 3, we demonstrate that if the pseudo labels are biased, the entire-space CVR estimator will also be biased. We propose a conditional entire-space CVR teacher to generate unbiased pseudo labels. The na\"ive entire-space CVR teacher model CVR as $\mathbb{P}(r_{u,i}=1)$, which is biased since users in the click space have a higher chance of conversion \cite{schnabel2016recommendations}:
\begin{equation}
    \mathbb{E}_{\mathcal{D}}[\mathbb{P}(r_{u,i}=1)] \neq \mathbb{E}_{\mathcal{D}}[\mathbb{P}(r_{u,i}=1|o_{u,i}=1)].
\end{equation}

\begin{figure*}[htb]
\centering
\includegraphics[width=1.6\columnwidth]{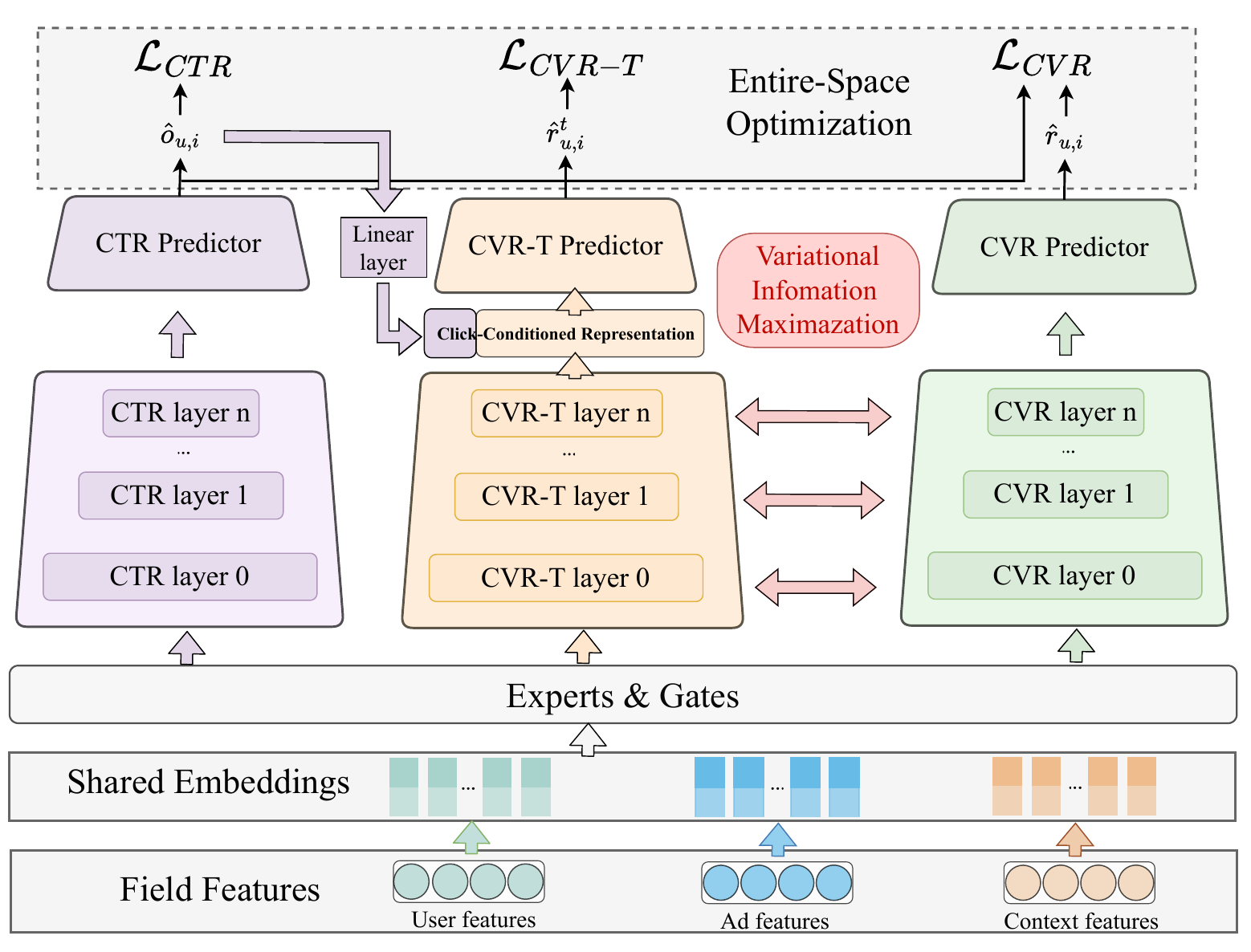}

\caption{Architecture of EVI. The EVI consists of CTR, CVR-T (teacher) and CVR estimator (student) with shared embedding layer. Following the Experts \& Gates, three three multilayer perceptrons serve as representation learners and predictors. CVR-T (teacher) produces unbiased pseudo conversion labels based on click-conditioned representations. We maximize the variational information between CVR-T and CVR representation learner to tranfer entire-space conversion knowledge to CVR estimator.}
\label{Framework}
\end{figure*}

The conditional entire-space CVR teacher model CVR as $\mathbb{P}(r_{u,i}=1|do(o_{u,i}=1)$.  The 'do' operator, referring to do-calculus, removes the dependency from click to conversion, ensuring unbiased results \cite{wang_escm2_2022}. In this paper, we adopt the click propensity conditioning approach, which is similar to IPS \cite{zhang_large-scale_2020}. Specifically, we train a CTR estimator to predict the click propensity, which can be used to represent all the confounders \cite{rubin2005causal}. We then project the click propensity through a linear layer and use the outer product of this projection and the CVR-T representation learner's output to generate the click-conditioned representation, as shown in Figure \ref{Framework}. Based on the click-conditioned representation, we can model the teacher's CVR as $\mathbb{P}(r_{u,i}=1|\hat{O}_{u,i}=\hat{o}_{u,i})$.
The unbiasedness of the conditional entire-space CVR teacher and EVI CVR estimator are proven in Theorem 2 and 3.

\begin{theorem}
The conditional entire-space CVR teachers are unbiased, provided that $\hat{o}_{u,i}$ is accurate, i.e.,
\begin{equation}
\mathbb{E}_{\mathcal{D}}[\mathbb{P}(r_{u,i}=1| \hat{O}_{u,i}=\hat{o}_{u,i}] = \mathbb{E}_{\mathcal{D}}[\mathbb{P}(r_{u,i}=1|o_{u,i}=1)]. \notag
\end{equation}
\end{theorem}
\begin{proof}
    See Appendix B.2.
\end{proof}

\begin{theorem}
    The EVI CVR estimator is unbiased when the click propensity $\hat{o}_{u,i}$ and pseudo conversion labels of unclicked samples $r^*_{u,i}$ are accurate, i.e., \begin{equation}|\mathbb{E}_{\mathcal{D}}\left\lbrack\mathcal{L}_{\mathrm{CVR}}^{\mathrm{EVI}}\right\rbrack-\mathcal{L}_{\mathrm{CVR}}^{ideal}|=0. \notag
    \end{equation}
\end{theorem}
\begin{proof}
    See Appendix B.3.
\end{proof}

\begin{table*}[h]
\centering
\begin{tabular}{ccccccccl} 
\toprule
Methods        & VIE           & CECT          & Ali-CCP                  & AE-FR           & AE-ES           & AE-US           & AE-NL           & Industrial       \\ 
\hline
EVI w/o VIE~CECT & $\times$     & $\times$     & 0.6702                   & 0.7777          & 0.7943          & 0.7888          & 0.7675          & 0.8218           \\
EVI w/o VIE     & $\times$     & $\checkmark$ & 0.6776                   & 0.7971          & 0.8075          & 0.8015          & 0.7717          & 0.8254           \\
EVI            & $\checkmark$ & $\checkmark$ & \textbf{\textbf{0.6802}} & \textbf{0.8008} & \textbf{0.8140} & \textbf{0.8057} & \textbf{0.7741} & \textbf{0.8292}  \\
\bottomrule
\end{tabular}
\caption{Ablation study of EVI. Abbreviations: VIE (variational information exploitation) and CECT (conditional entire-space CVR teacher). Top CVR results are in \textbf{bold}.}
\label{Ablation CVR}
\end{table*}

\subsection{Variational Information Exploitation}

Existing entire-space methods use logits distillation to learn the non-click space information. However, the information carried in the logits (pseudo labels) is limited and easily biased.

From an information theory perspective, maintaining high mutual information between the teacher and student network layers can improve the transfer of non-click space information 
\cite{ahn2019variational}. To address the computational difficulty of mutual information, we enhance non-click space exploitation based on a variational information maximization scheme \cite{barber2004algorithm, ahn2019variational}, which maximizes the variational lower bound instead. Passing the feature embeddings $\mathbf{x}_{u,i}$ through the networks in Figures \ref{Framework} creates $K$ pairs of random variables $\{(\mathbf{t}^{(k)},\mathbf{s}^{(k)})\}_{k=1}^K$. 
These pairs represent the output of the learner layers for CVR-T (teacher) and CVR (student). The variational lower bound for the mutual information $I(\mathbf{t};\mathbf{s})$ of each layer pair is defined as follows:
\begin{equation} 
\begin{aligned}
&I(\mathbf{t};\mathbf{s})=H(\mathbf{t})-H(\mathbf{t}|\mathbf{s}) \\
&=H(\mathbf{t})+\mathbb{E}_{\mathbf{t},\mathbf{s}}[\log p(\mathbf{t}|\mathbf{s})] \\
&=H(\mathbf{t})+\mathbb{E}_{\mathbf{t},\mathbf{s}}[\log q(\mathbf{t}|\mathbf{s})]+\mathbb{E}_{\mathbf{s}}[D_{\mathrm{KL}}(p(\mathbf{t}|\mathbf{s})||q(\mathbf{t}|\mathbf{s}))] \\
&\geq H(\mathbf{t})+\mathbb{E}_{\mathbf{t},\mathbf{s}}[\log q(\mathbf{t}|\mathbf{s})],
\end{aligned}
\end{equation}
where entropy $H(\mathbf{t})$ and the conditional entropy $H(\mathbf{t}|\mathbf{s})$
are calculated from the joint distribution $p(\mathbf{t}, \mathbf{s})$. $D_{KL}$ is Kullback-Leiber divergence \cite{goldberger2003efficient}. $q(\mathbf{t}, \mathbf{s})$ is a variational distribution that approximates $p(\mathbf{t}, \mathbf{s})$. In this work, we employ Gaussian distribution with mean $\boldsymbol{\mu}$ and $\boldsymbol{\rho}$ as the variational distribution $q(\mathbf{t}, \mathbf{s})$. We parameterize the $\boldsymbol{\mu}$ with a linear layer of parameters $\theta_{\mathbf{s}}$, i.e., $\boldsymbol{\mu} = f(\mathbf{s};\theta_{\mathbf{s}})$. We set \(\boldsymbol{\rho}\) as trainable parameters that pass through a softplus function \cite{zheng2015improving}, ensuring their positivity.
Ignoring the constant $H(\mathbf{t})$, the variational information exploitation (VIE) loss is defined as:
\begin{equation}
    \begin{aligned} & \mathcal{L}_{\mathrm{VIE}}^{\mathrm{EVI}}=\sum_{n=1}^{N}\log\sigma_{n}+\frac{(t_n-\mu_n(\boldsymbol{s}))^2}{2\sigma_n^2}\end{aligned},
\end{equation}
where $t_n$, $\mu_{n}$ and $\rho_{n}$ indicates the n-th entry of the vector $\mathbf{t}$, $\boldsymbol{\mu}$ and $\boldsymbol{\rho}$.

\begin{figure*}[h]
\centering
\subfloat[Logloss on Ali-CCP]
{
\includegraphics[width=0.6\columnwidth]{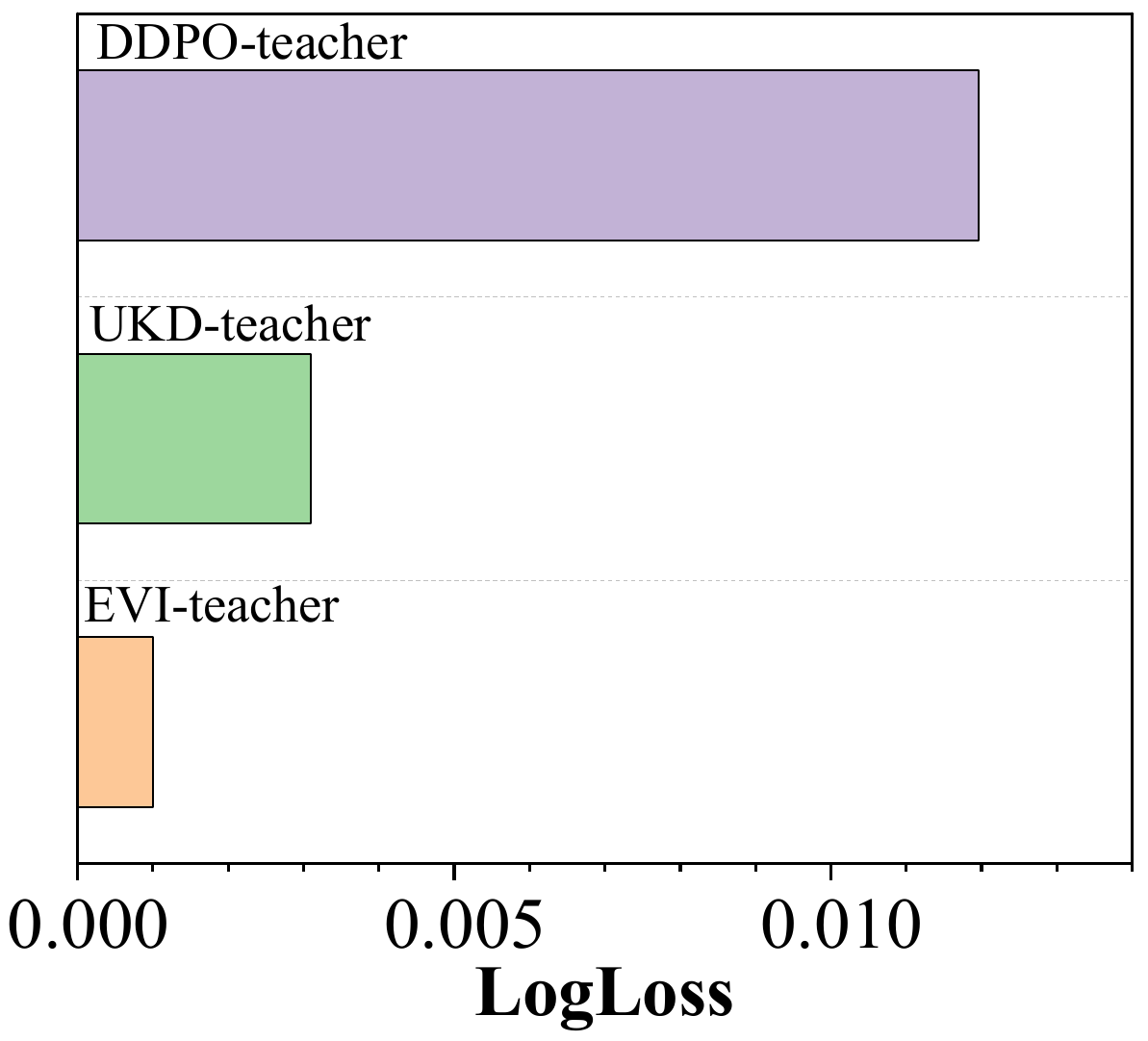}
}
\subfloat[Logloss on AE-FR]
{
\includegraphics[width=0.6\columnwidth]{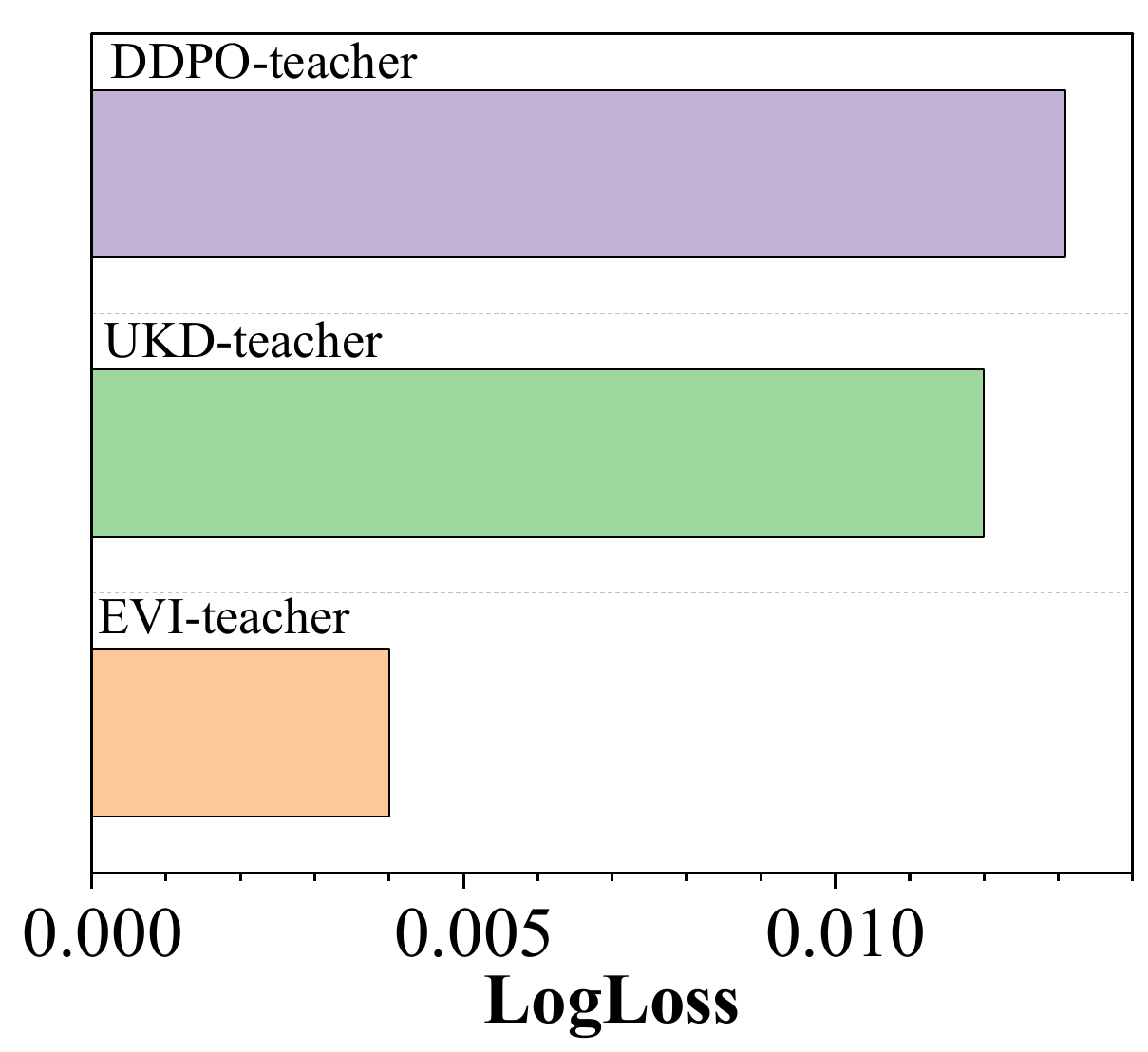}
}
\subfloat[Logloss on Industrial]
{
\includegraphics[width=0.6\columnwidth]{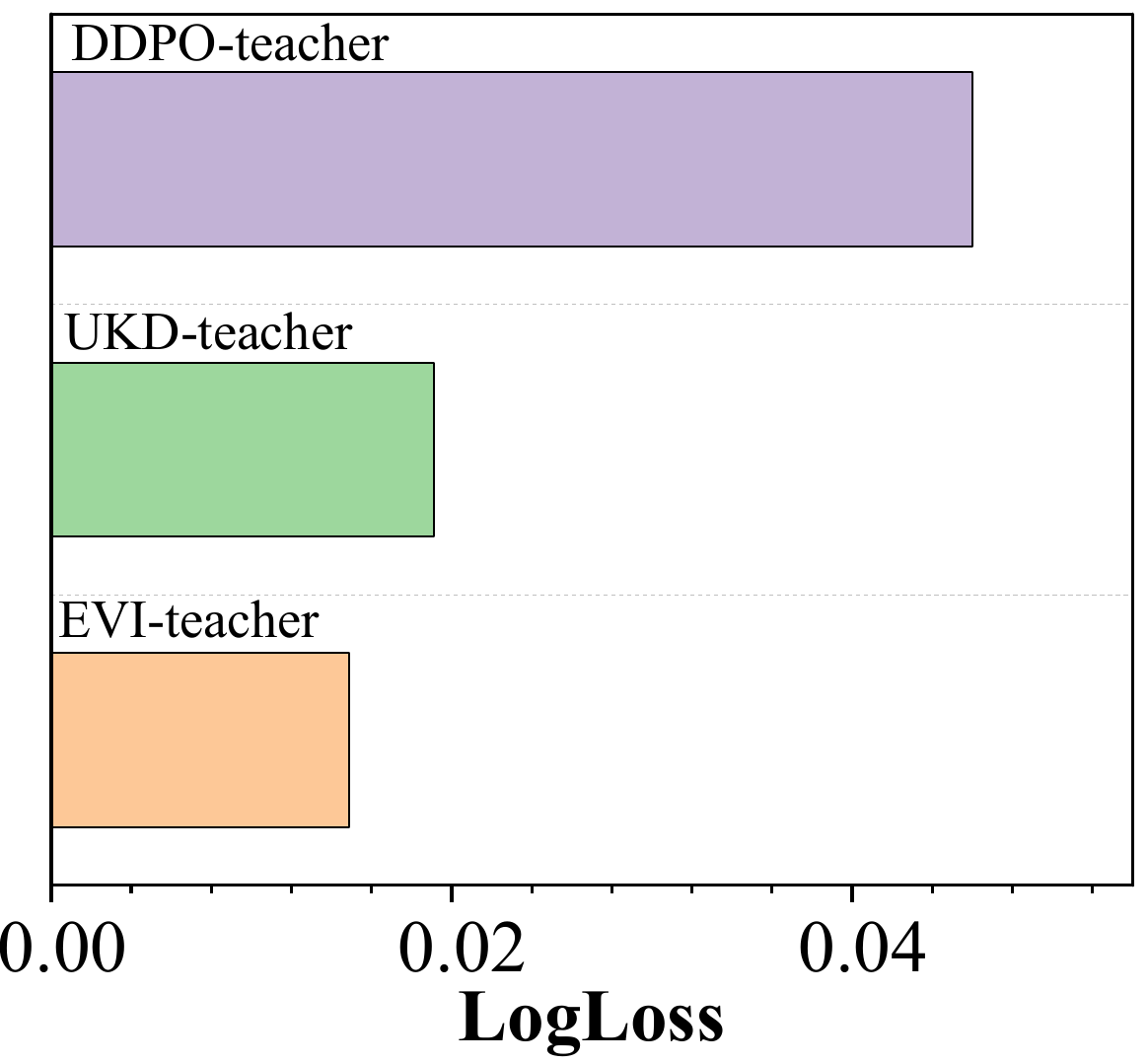}
}
\newline
\centering
\subfloat[Mean Bias on Ali-CCP]
{
\includegraphics[width=0.6\columnwidth]{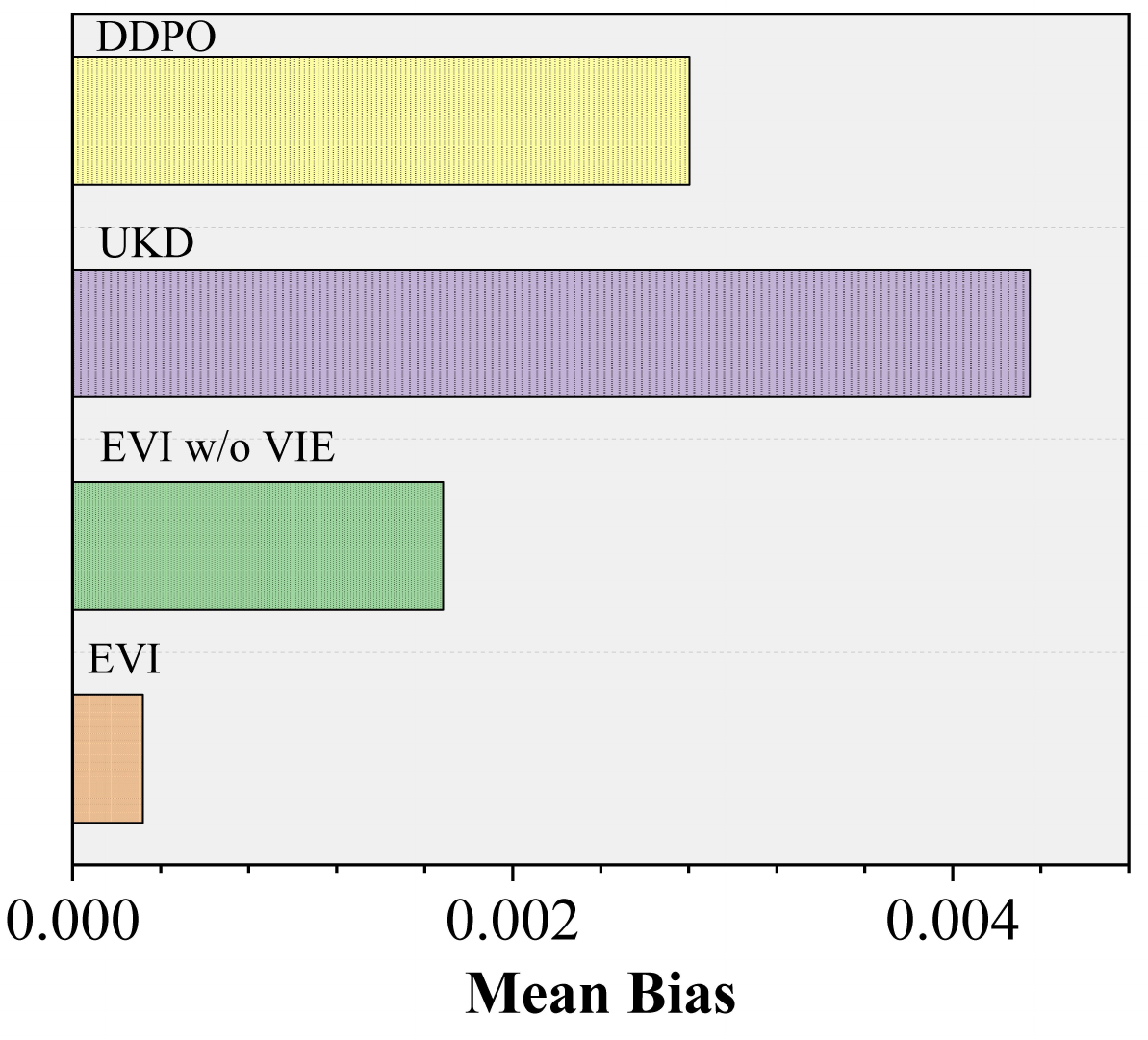}
}
\subfloat[Mean Bias on AE-FR]
{
\includegraphics[width=0.6\columnwidth]{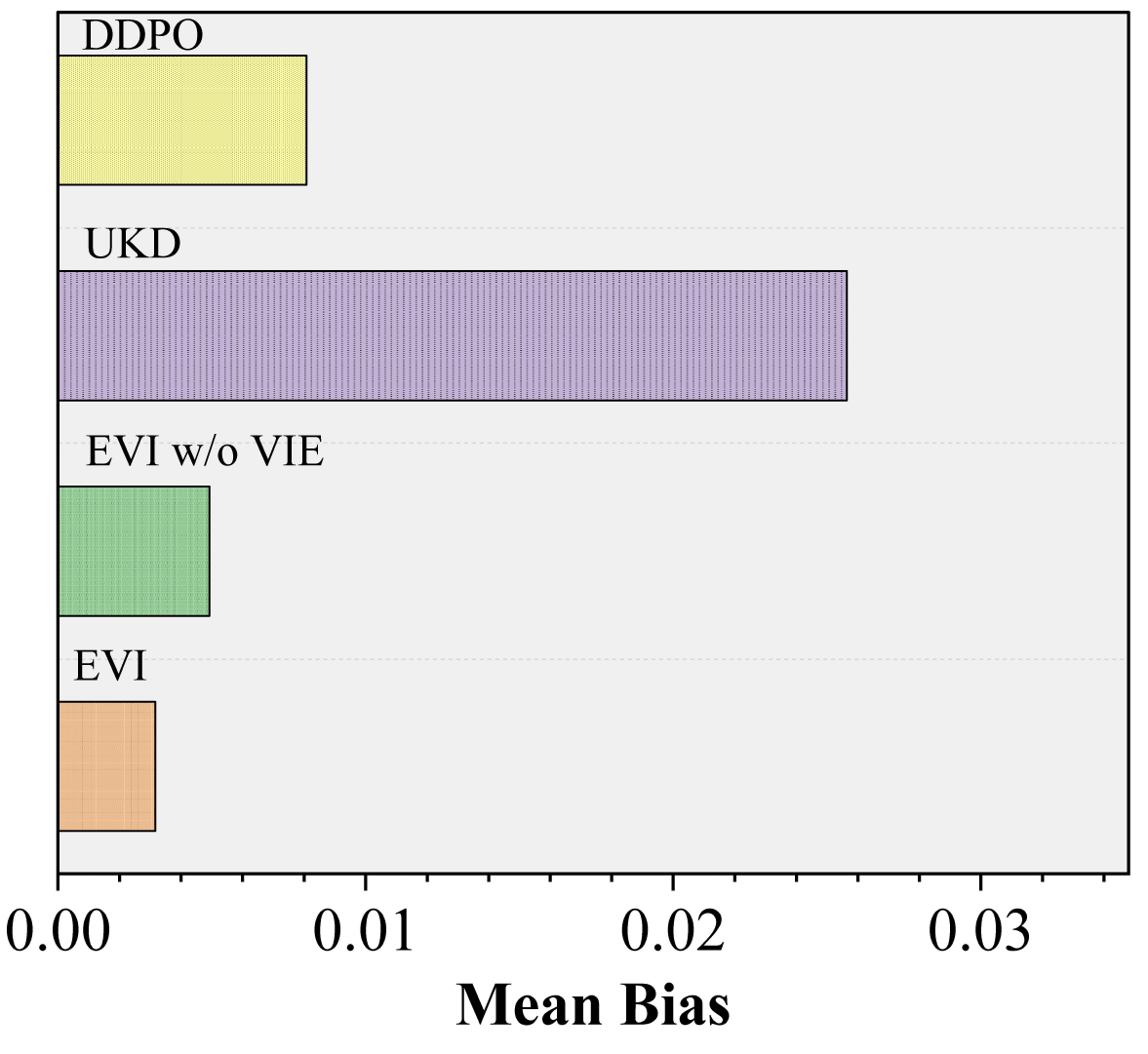}
}
\subfloat[Mean Bias on Industrial]
{
\includegraphics[width=0.6\columnwidth]{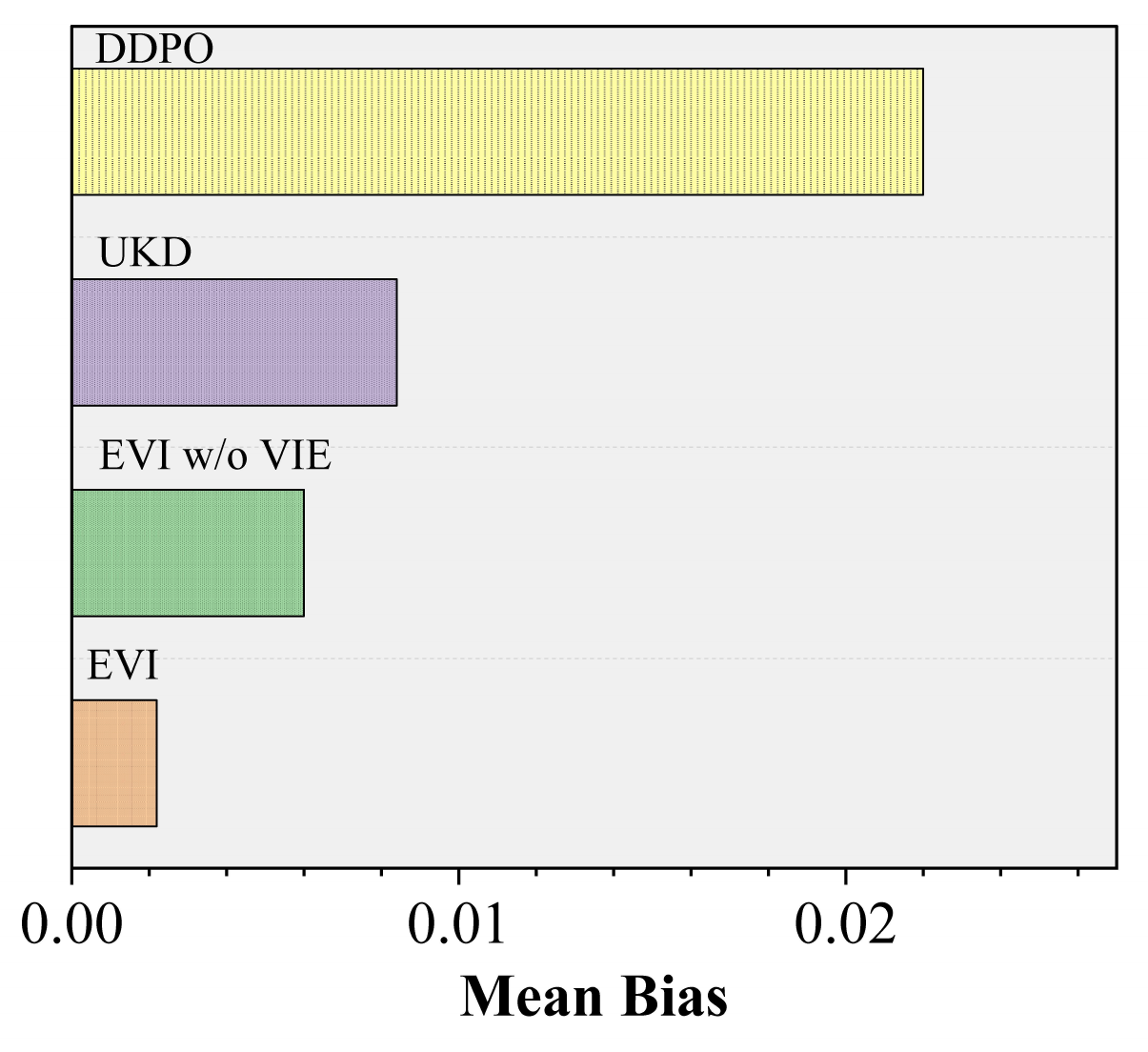}
}
\caption{The teachers' logloss on non-click space and the students' CVR mean bias. EVI w/o VIE means the EVI excludes variational information exploitation.}
\label{logloss and bias}
\end{figure*}

\subsection{Entire-space CVR Optimization}

In EVI, we optimize CTR, CVR-T, CVR and CTCVR in a single model with embedding sharing. The CTR and CTCVR loss are defined as:

\begin{align}\mathcal{L}_{\mathrm{CTR}}^{\mathrm{EVI}}=\frac{1}{|\mathcal{D}|}\sum_{(u,i)\in\mathcal{D}}\delta\left(o_{u,i},\hat{o}_{u,i}^{}\right),\end{align}
\begin{align}\mathcal{L}_{\mathrm{CTCVR}}^{\mathrm{EVI}}=\frac{1}{|\mathcal{D}|}\sum_{(u,i)\in\mathcal{D}}\delta\left(o_{u,i}*r_{u,i},\hat{o}_{u,i}^{}*\hat{r}_{u,i}\right),\end{align}
where $\delta$ denotes the binary cross-entropy loss function. We optimize the CVR teacher model over the entire space, conditioned on the click propensity $\hat{o}_{u,i}$.
the CVR-T loss for EVI, $\mathcal{L}_{\mathrm{CVR-T}}^{\mathrm{EVI}}$, is defined as follows:

\begin{align} 
& \mathcal{L}_{\mathrm{CVR-T}}^{\mathrm{EVI}}= \frac{1}{|\mathcal{D}|}\sum_{(u,i)\in\mathcal{D}}\delta(r_{u,i},\hat{r}_{u,i})|\mathbf{x}_{u,i},\hat{o}_{u,i},
\end{align}

For the CVR task, we optimize the loss function over the entire exposure space \(\mathcal{D}\) and employ inverse click propensity and non-click propensity to ensure the unbiasedness (see Theorem 3):

\begin{align}  \mathcal{L}_{\mathrm{CVR}}^{\mathrm{EVI}}&=\frac{1}{2|\mathcal{D}|}\sum_{(u,i)\in\mathcal{O}}\frac{o_{u,i}\delta(r_{u,i},\hat{r}_{u,i})}{\hat{o}_{u,i}}\notag \\  & +\frac{1}{2|\mathcal{D}|}\sum_{(u,i)\in\mathcal{N}}\frac{\left(1-o_{u,i})\delta(r_{u,i}^{*},\hat{r}_{u,i}\right)}{1-\hat{o}_{u,i}}\end{align}

So the overall loss are defined as:
\begin{align}
\mathcal{L}_{\mathrm{EVI}} &=\lambda_{c}\mathcal{L}_{\mathrm{CTR}}^{\mathrm{EVI}}+\lambda_{t}\mathcal{L}_{\mathrm{CVR-T}}^{\mathrm{EVI}}+\lambda_{r}\mathcal{L}_{CVR}^{\mathrm{EVI}}\notag \\ &+\lambda_{i}\mathcal{L}_{\mathrm{VIE}}^{\mathrm{EVI}}+\lambda_{g}\mathcal{L}_{\mathrm{CTCVR}}^{\mathrm{EVI}},
\end{align}
where $\lambda_{c}$, $\lambda_{t}$, $\lambda_{r}$, $\lambda_{i}$ and $\lambda_{g}$ are trade-offs for each tasks.

\section{Experiments}

We conduct experiments using five large-scale public datasets and one private industrial dataset to validate the effectiveness of EVI. We compare EVI with 8 CVR estimation baselines. These baselines include both causal and non-causal methods, as well as entire-space and partial-space models. See Appendix A.1 for dataset information, Appendix A.2 for baseline details, and Appendix A.3 for implementation details.

\begin{table*}[h]
\centering
\resizebox{\textwidth}{!}{
\begin{tabular}{ccccccccccccc} 
\toprule
\multirow{2}{*}{Methods}                           & \multicolumn{2}{c}{Ali-CCP}             & \multicolumn{2}{c}{AE-FR}               & \multicolumn{2}{c}{AE-ES}               & \multicolumn{2}{c}{AE-US}               & \multicolumn{2}{c}{AE-NL}               & \multicolumn{2}{c}{Industrial}           \\
                                                   & AUC                & NLL                & AUC                & NLL                & AUC                & NLL                & AUC                & NLL                & AUC                & NLL                & AUC                & NLL                 \\ 
\hline
ESMM \cite{ma_entire_2018}                         & 0.6703             & 0.0331             & 0.7485             & 0.1148             & 0.8011             & 0.0983             & 0.7241             & 0.1055             & 0.6931             & 0.1463             & 0.8127             & 0.1694              \\
ESCM\textsuperscript{2}-IPW~\cite{wang_escm2_2022} & 0.6622             & 0.0335             & \underline{0.7865} & \underline{0.1116} & 0.7905             & 0.0962             & 0.7885             & 0.1086             & 0.7554             & 0.1798             & 0.8208             & 0.1644              \\
ESCM\textsuperscript{2}-DR~\cite{wang_escm2_2022}  & 0.6494             & 0.0610             & 0.7861             & 0.1269             & 0.8004             & 0.1202             & 0.7887             & 0.1086             & 0.7651             & 0.1755             & 0.8203             & 0.1649              \\
~MRDR-GPL~\cite{zhou2023generalized}               & 0.6195             & 0.1076             & 0.7600             & 0.1972             & 0.7749             & 0.1707             & 0.7624             & 0.1652             & 0.7452             & 0.2312             & 0.8173             & 0.2088              \\
DR-V2~\cite{li2023propensity}                      & 0.6093             & 0.0580             & 0.7656             & 0.1192             & 0.7846             & 0.1140             & 0.7591             & 0.1025             & 0.7170             & 0.1592             & 0.8188             & 0.2096              \\
UKD \cite{xu2022ukd}                               & 0.6594             & 0.0375             & 0.7699             & 0.1156             & \underline{0.8032} & \underline{0.0951} & 0.7790             & 0.1010             & 0.7281             & 0.1448             & \underline{0.8211} & \underline{0.1651}  \\
DCMT~\cite{zhu2023dcmt}                            & 0.6705             & 0.0333             & 0.7718             & 0.1146             & 0.7931             & 0.0952             & \underline{0.7889} & \underline{0.0997} & 0.7339             & \underline{0.1425} & 0.8192             & 0.1684              \\
DDPO~\cite{su2024ddpo}                             & \underline{0.6711} & \underline{0.0330} & 0.7751             & 0.1194             & 0.7913             & 0.0976             & 0.7712             & 0.1062             & \underline{0.7685} & 0.1449             & 0.8202             & 0.1687              \\
EVI                                                & \textbf{0.6802}    & \textbf{0.0321}    & \textbf{0.8008}    & \textbf{0.1037}    & \textbf{0.8136}    & \textbf{0.0939}    & \textbf{0.8057}    & \textbf{0.0976}    & \textbf{0.7741}    & \textbf{0.1401}    & \textbf{0.8292}    & \textbf{0.1621}     \\
\bottomrule
\end{tabular}
}
\caption{CVR results for six datasets. \underline{Underlined values }indicate the best baseline results, and \textbf{bold values} show the top results.}
\label{CVR Results}
\end{table*}

\subsection{Evaluation Metrics}
We use the area under the ROC curve (AUC) \cite{fawcett2006introduction} and Logloss (NLL) \cite{xu2022ukd} as evaluation metrics. Each experiment is conducted five times, and we report the average results. The significance of any improvement or decrease in performance is judged using t-test.
\begin{figure*}[htbp]
\centering
\subfloat[Ali-CCP]
{
\includegraphics[width=0.4\columnwidth]{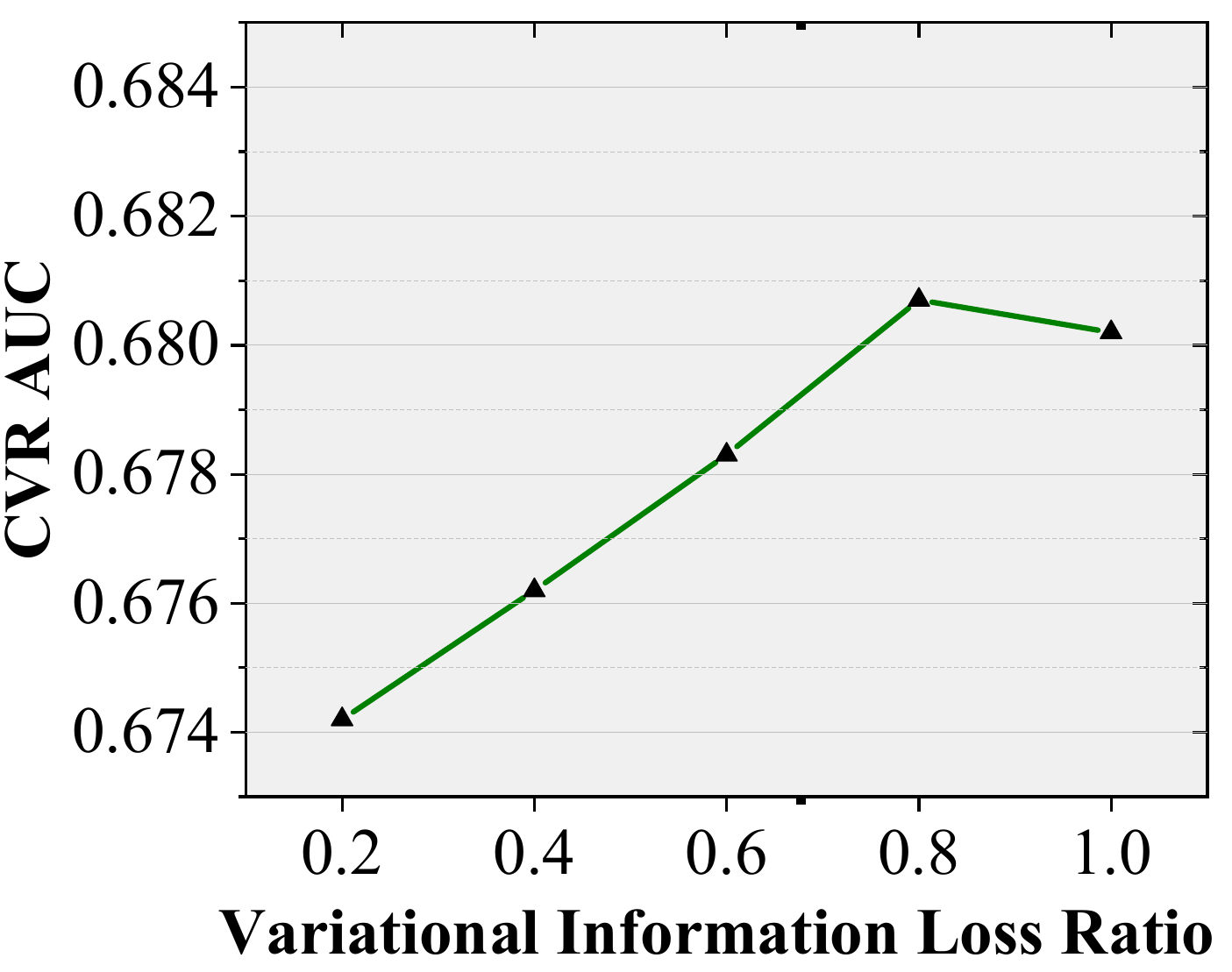}
}
\subfloat[AE-FR]
{
\includegraphics[width=0.4\columnwidth]{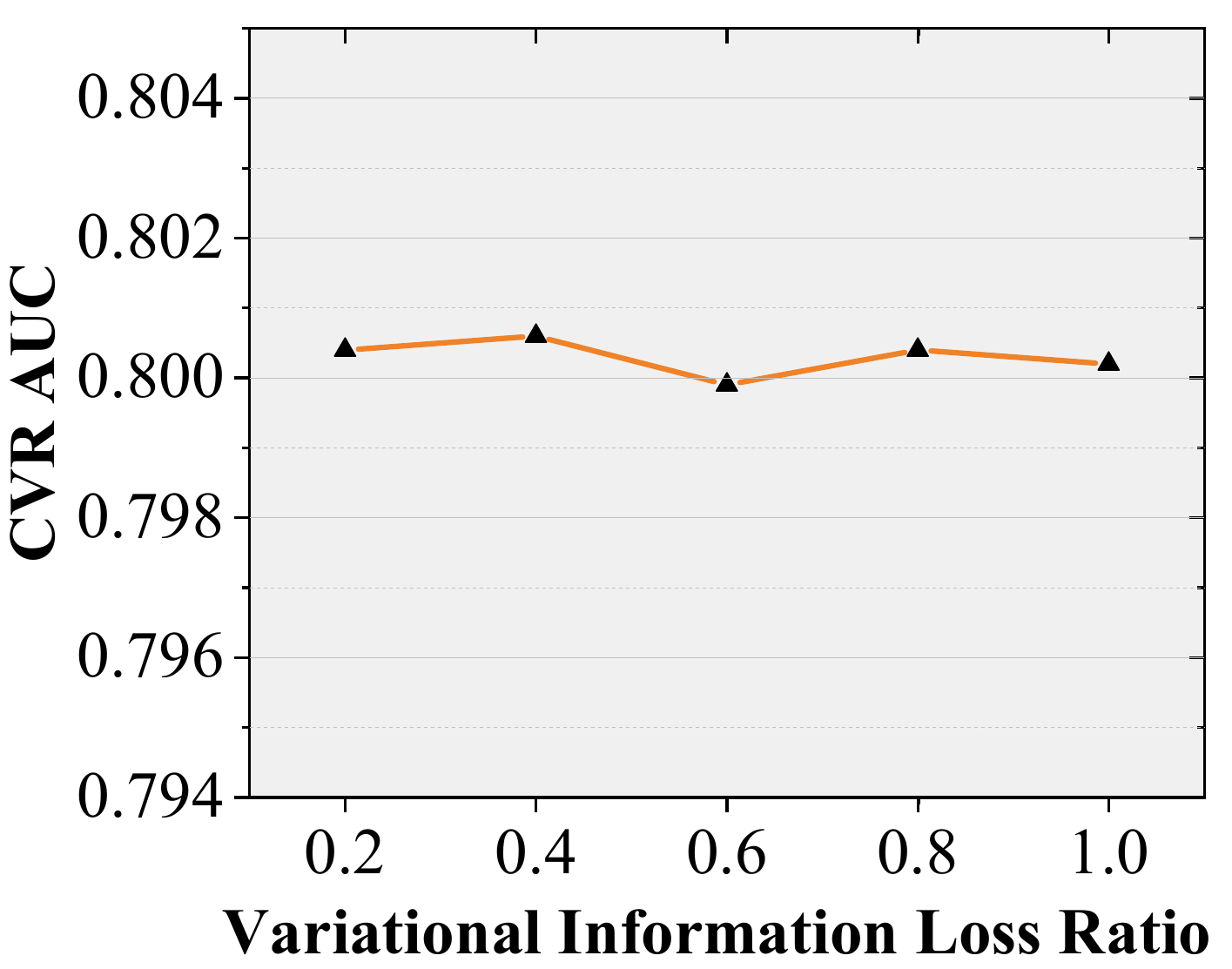}
}
\subfloat[AE-ES]
{
\includegraphics[width=0.4\columnwidth]{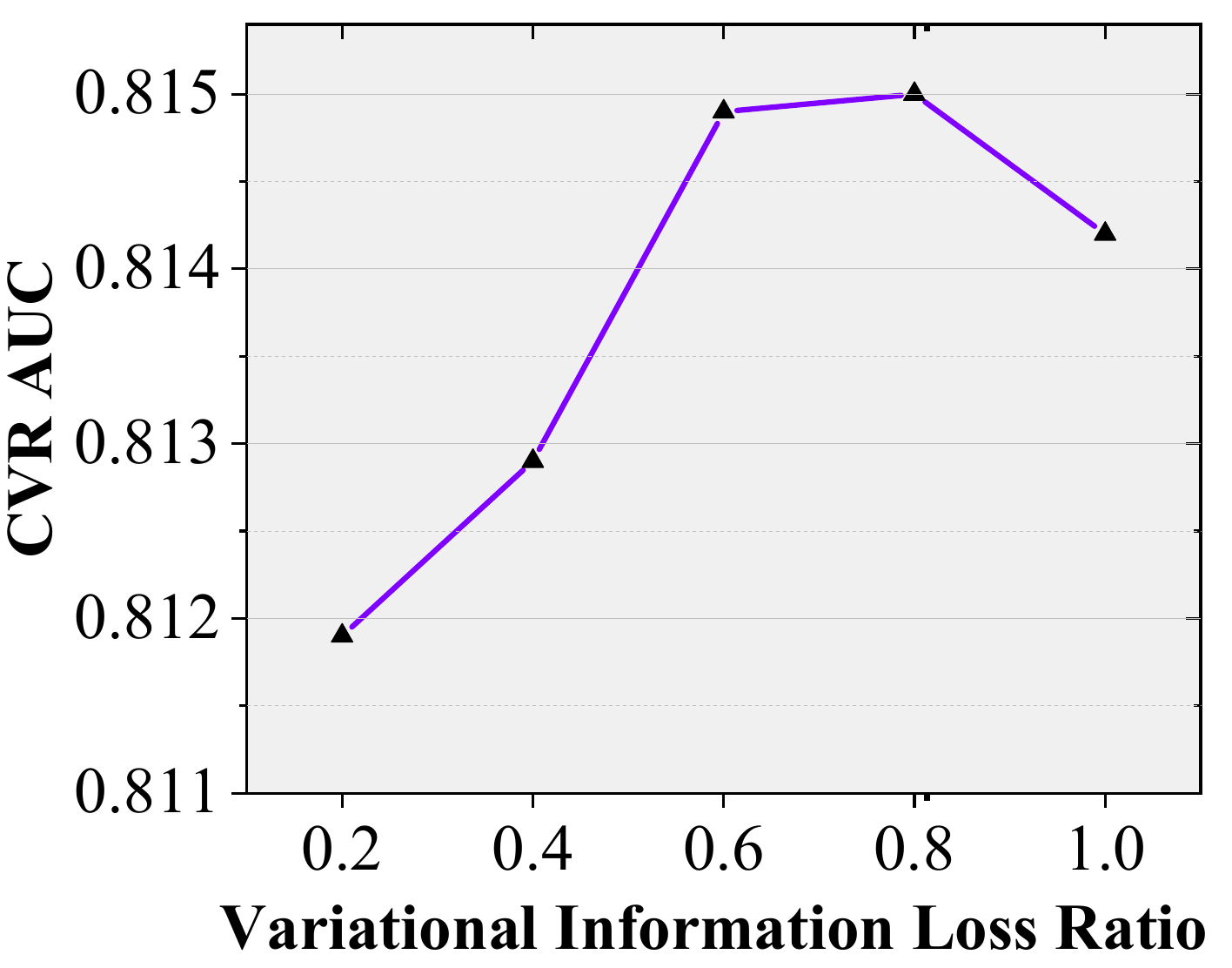}
}
\subfloat[AE-NL]
{
\includegraphics[width=0.4\columnwidth]{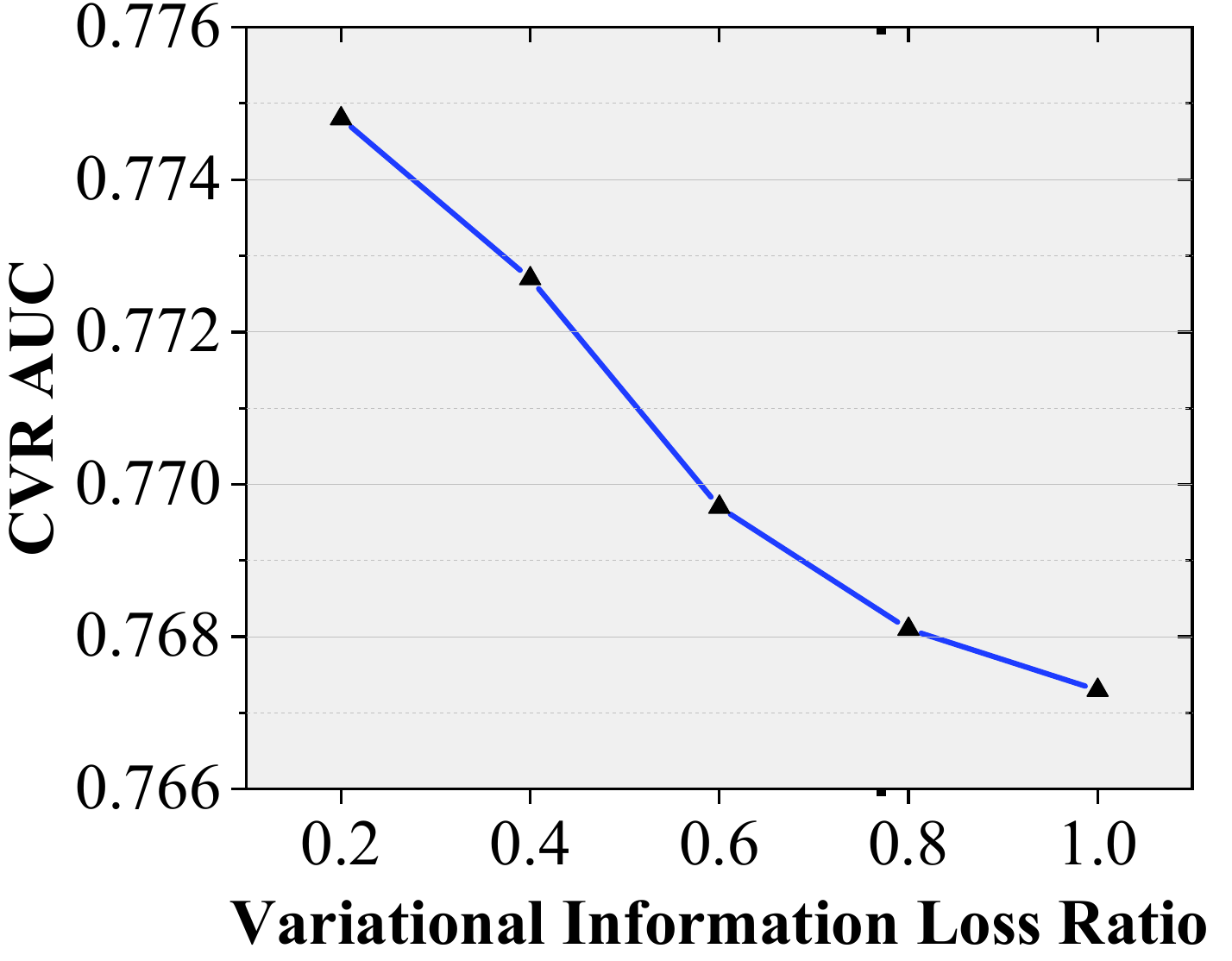}
}
\subfloat[Industrial]
{
\includegraphics[width=0.4\columnwidth]{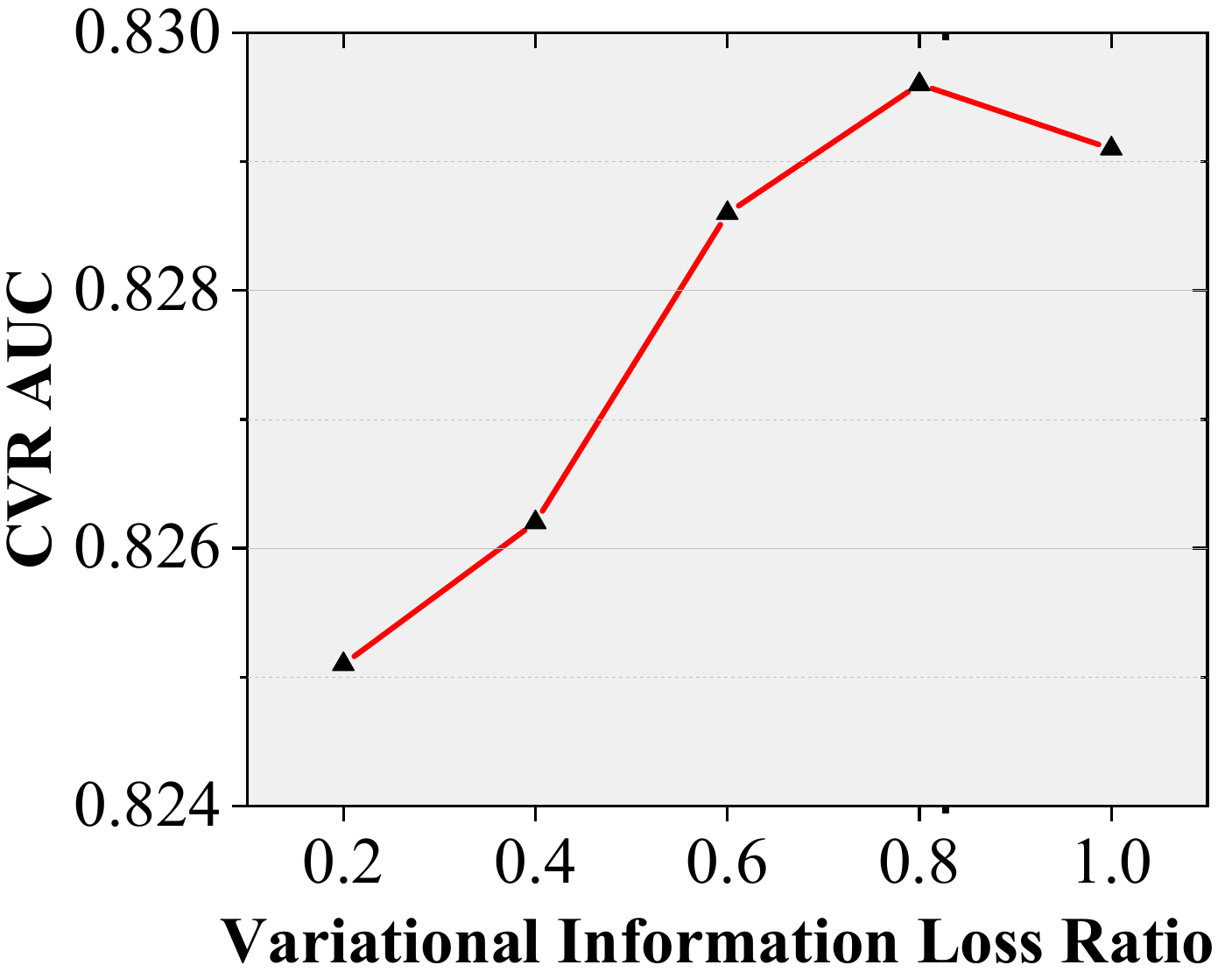}
}
\newline
\subfloat[Ali-CCP]
{
\includegraphics[width=0.4\columnwidth]{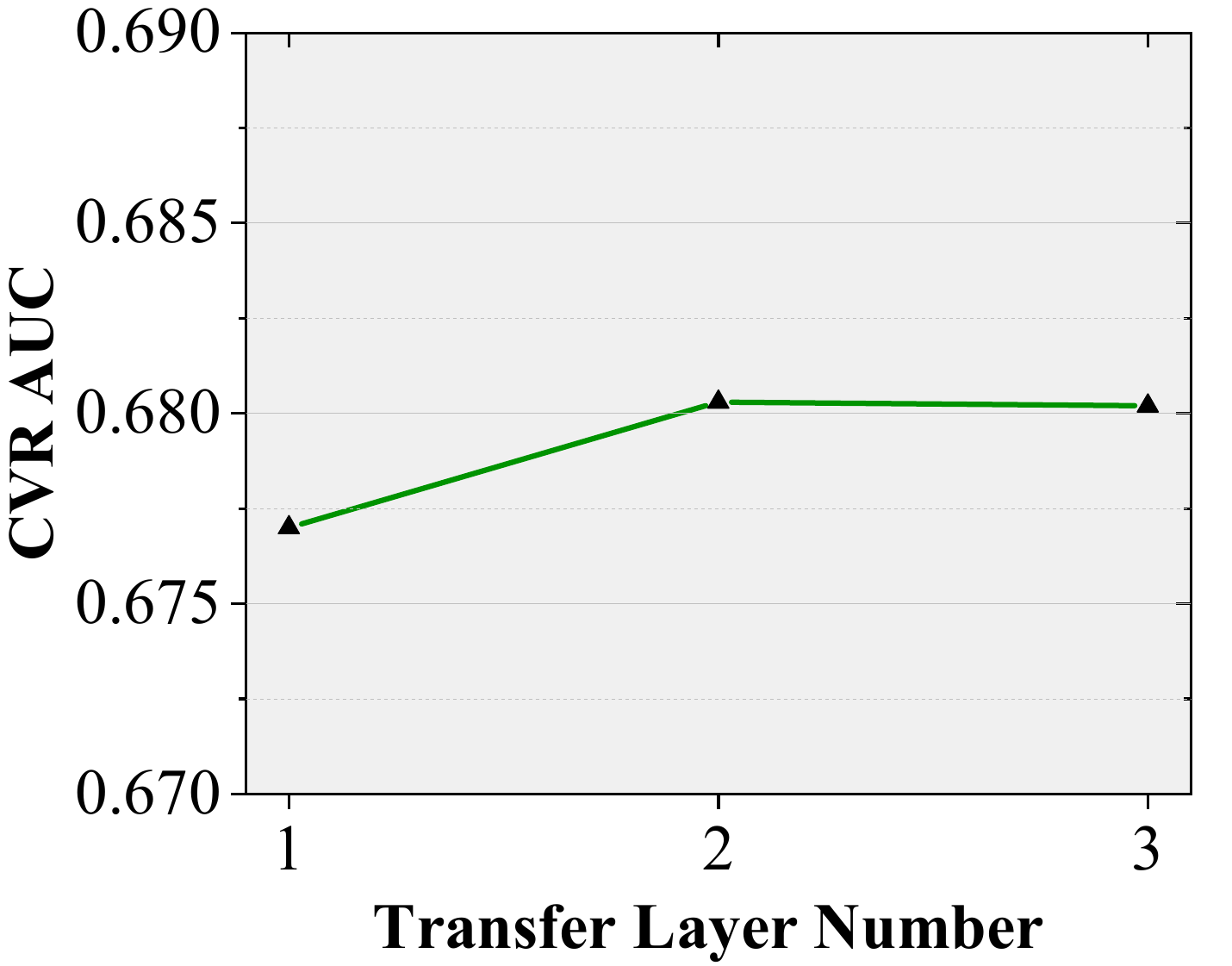}
}
\subfloat[AE-FR]
{
\includegraphics[width=0.4\columnwidth]{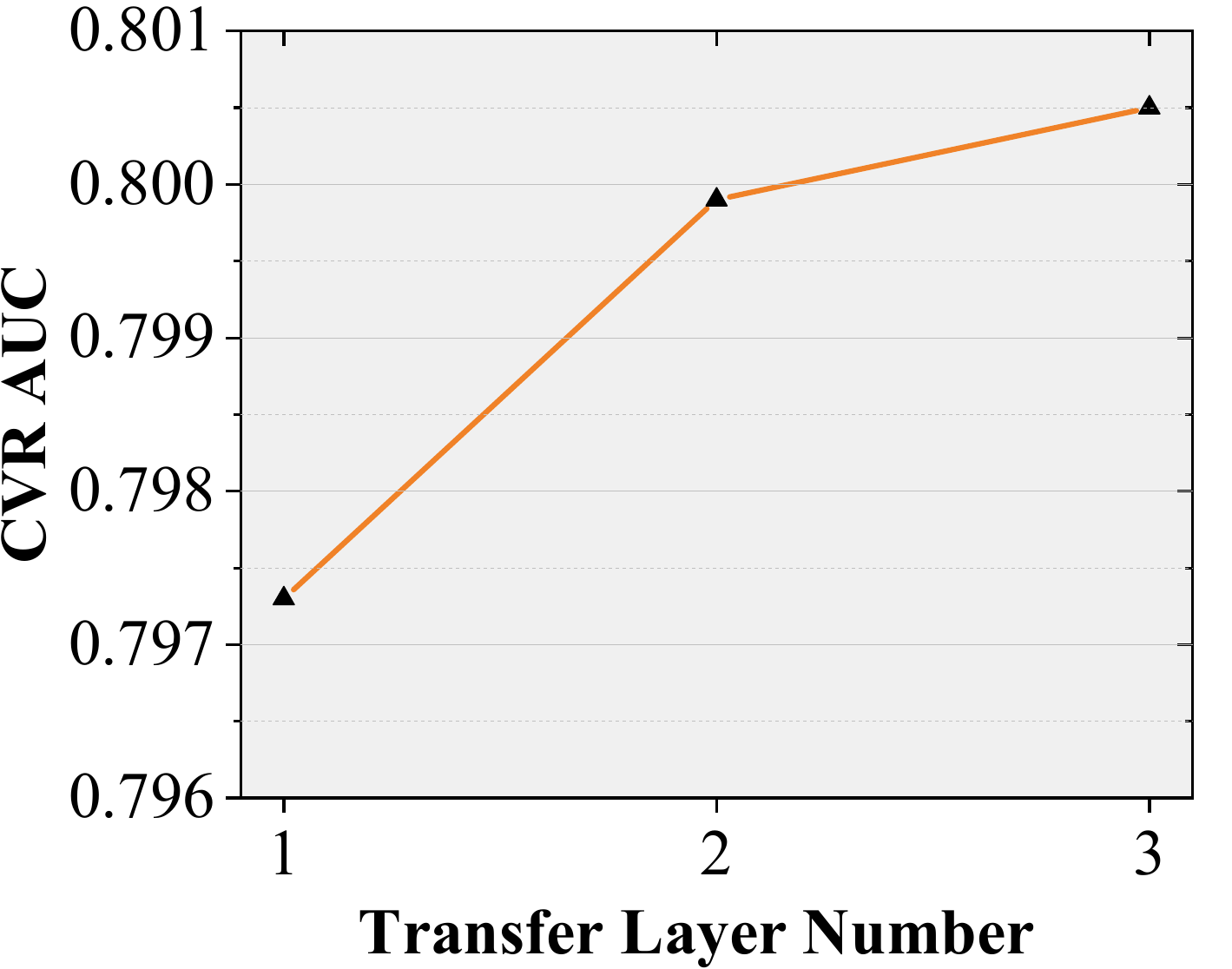}
}
\subfloat[AE-ES]
{
\includegraphics[width=0.4\columnwidth]{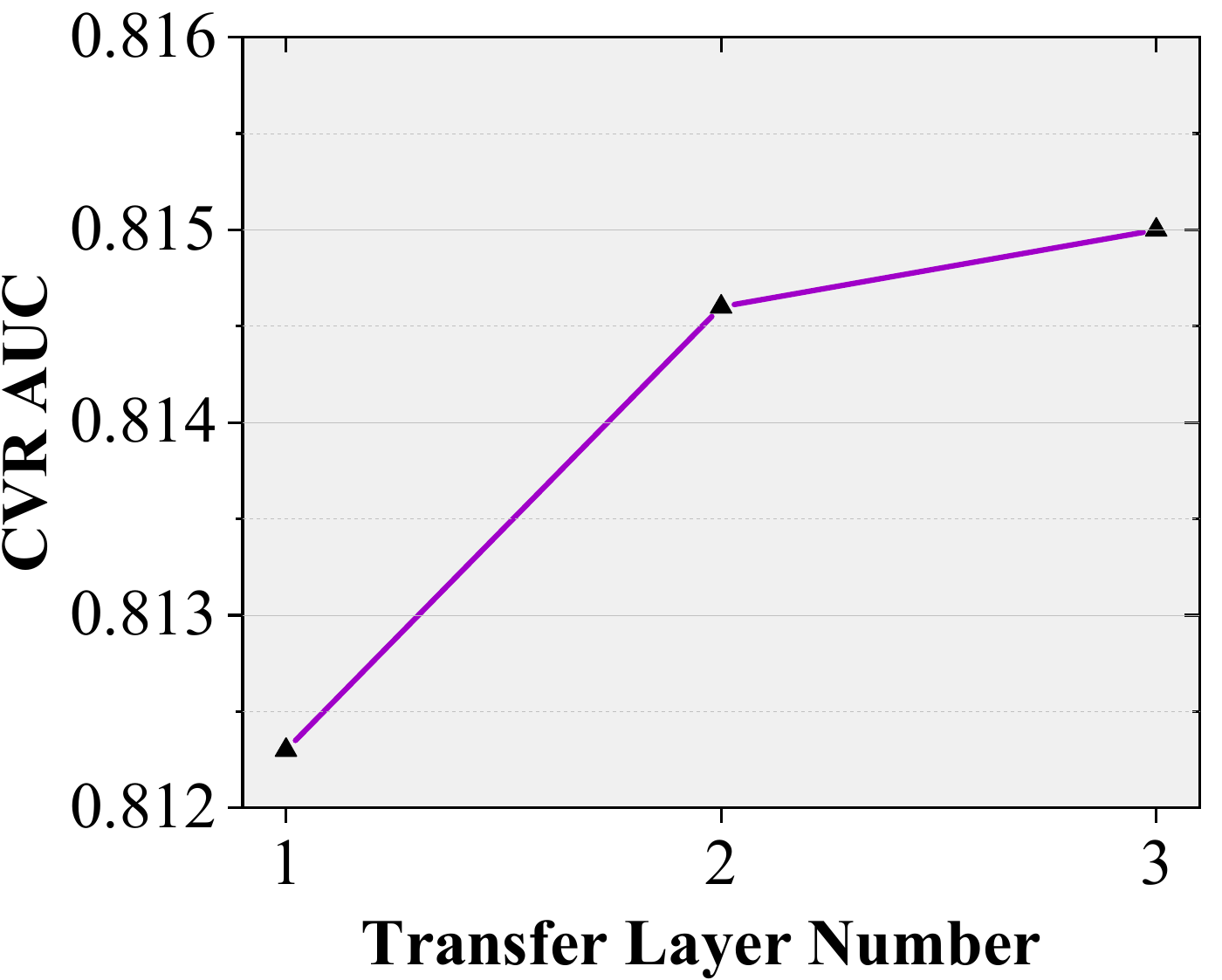}
}
\subfloat[AE-NL]
{
\includegraphics[width=0.4\columnwidth]{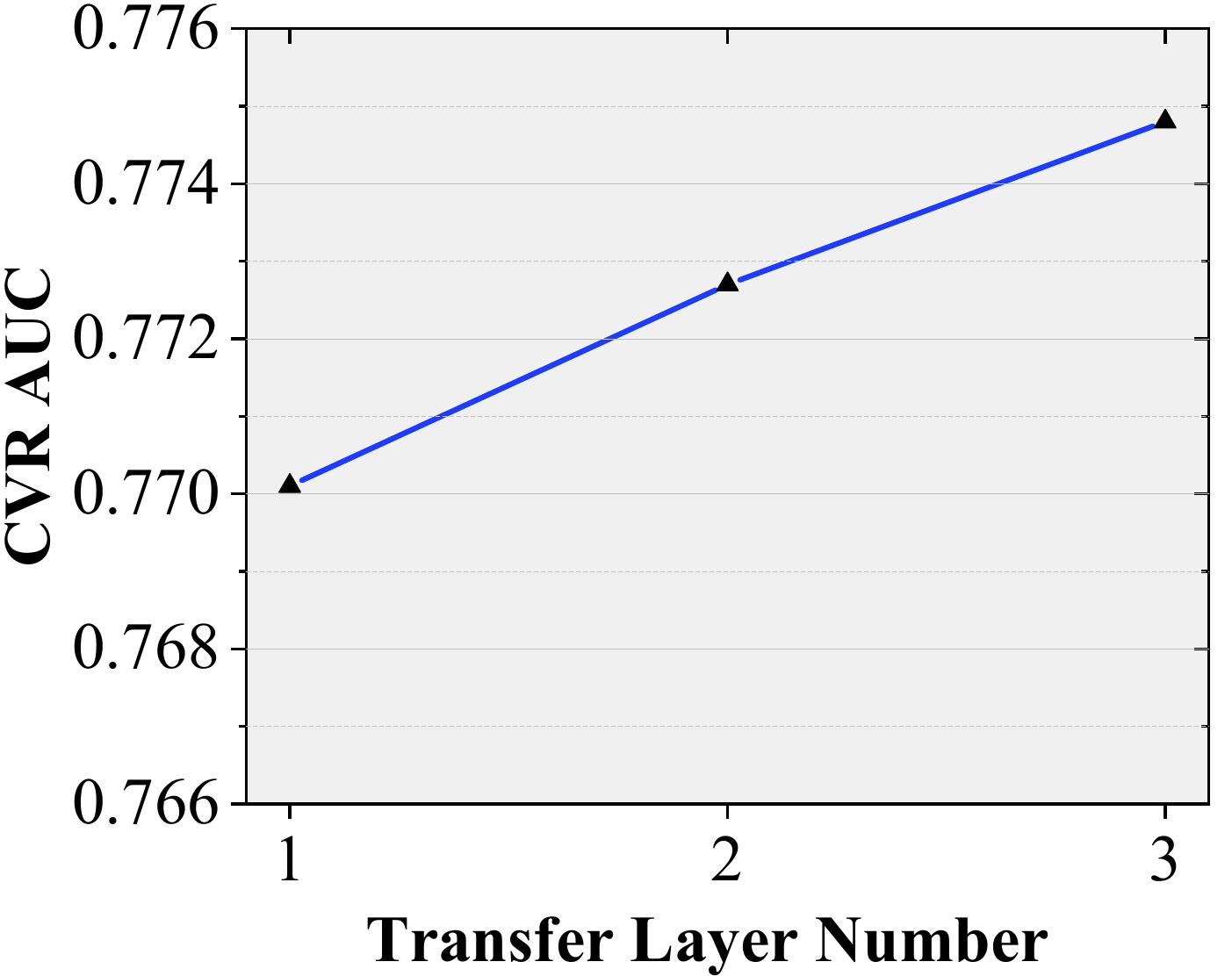}
}
\subfloat[Industrial]
{
\includegraphics[width=0.4\columnwidth]{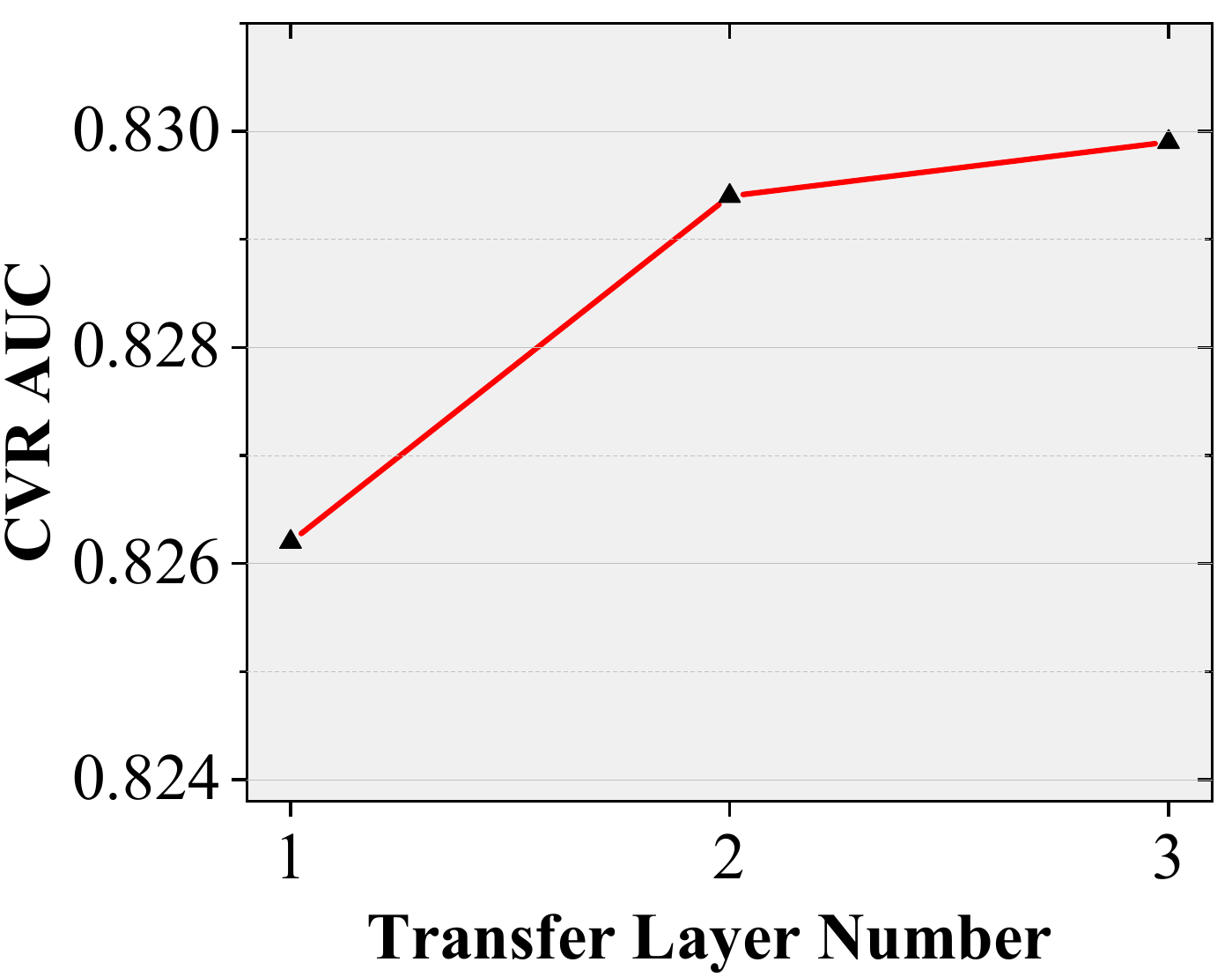}
}

\caption{Effects of varying VIE loss ratio and the number of transfer layers on four public and one industrial datasets.}
\label{Parameter sensitivity}
\end{figure*}

\subsection{CVR Estimation Performance Comparison}
Table \ref{CVR Results} presents a comparison of CVR task results between our EVI model and eight baselines, evaluated across both public and industrial datasets. Key observations can be drawn as:
\begin{itemize}
    \item The click-space causal methods demonstrate performance comparable to the ESMM model. The ESMM model initially introduced the entire-space CVR estimator. However, it performs poorly on the AE-FR, AE-US, AE-NL, and industrial dataset. This poor performance may be due to its inherent estimation bias \cite{wang_escm2_2022}. Applying causal inference to correct the inherent bias, click-space causal methods such as MRDR-GPL and DR-V2 outperform ESMM on AE-FR, AE-US, AE-NL, and Industrial datasets. However, they perform worse than ESMM on the Ali-CCP and AE-ES datasets. 
    \item Improved entire-space methods outperform both click-space causal methods and ESMM. Since entire-space exploitation and causal inference are not adversarial, enhanced entire-space methods combine causal inference and entire-space exploitation techniques. The experimental results across all datasets show that improved entire-space methods (EVI, ESCM$^2$-IPW, ESCM$^2$-DR, DCMT, UKD, DDPO, DCMT) outperform both click-space causal methods and ESMM. These findings are consistent with previous studies. Among the enhanced entire-space baselines, DDPO performs best in Ali-CCP and AE-NL, while UKD leads in AE-ES and Industrial datasets. ESCM$^2$-IPW and DCMT also excel in AE-FR and AE-US, respectively. Due to varying data characteristics, different methods show varied performance across datasets.
    \item Our EVI method consistently outperforms all baseline models in CVR estimation AUC across five datasets. It achieves the highest scores in Ali-CCP (0.6802), AE-FR (0.8008), AR-ES (0.8126), AE-US (0.8057), AE-NL (0.7741), and the industrial dataset (0.8292), with improvements of 1.3\%, 1.82\%, 1.34\%, 3.13\%, 4.84\%, and 0.99\% respectively.
    For NLL, EVI achieves the lowest scores in Ali-CCP (0.0321), AE-FR (0.1037), AR-ES (0.0939), AE-US (0.0976), AE-NL (0.1401), and industrial dataset (0.1621), reducing losses by 1.46\%, 1.89\%, 1.61\%, 2.17\%, and 2.77\% respectively. In general, EVI shows an average AUC improvement of 2.25\% and 2.78\% NLL reduction over the optimal baselines.
\end{itemize}

\subsection{Discussion on EVI}

In Section 3, through theoretical analysis, we found that biased pseudo labels transfer their bias to the entire-space CVR estimator. Unlike existing methods that use biased CVR teachers, we introduce an unbiased entire-space teacher trained on a click-conditioned representation. It is crucial to compare the bias of pseudo-labels produced by different CVR teachers using real datasets.

Specifically, we measure the logloss of the pseudo-labels in the non-click space provided by the UKD teacher, DDPO teacher, and EVI teachers. The logloss results conducted on the test sets of Ali-CCP, AE-FR, and the industrial dataset are shown in Figure \ref{logloss and bias}. The logloss for the DDPO-teacher is the highest across the three datasets. This is because the DDPO-teacher is a na\"ive CVR estimator trained on the click space, making it prone to overestimating the pCVR on the unclick space, consistent with previous studies \cite{wang_escm2_2022}. The logloss of the UKD is lower than that of the DDPO-teacher, possibly due to domain adaptation. Domain adaptation aligns the distributions of the click domain and the non-click domain, thus reducing overestimation. The EVT-teacher attains the lowest logloss, validating the theoretical analysis.
 
To validate how the pseudo labels and variational information exploitation affect the overall bias of the target CVR estimator, we measure the mean bias of the CVR estimate of the UKD, DDPO, EVI w/o VIE and EVI. The mean bias results, conducted on the test set of Ali-CCP, AE-FR, and the industrial dataset, are shown in Figure \ref{logloss and bias}. Although the pseudo-labels show more bias for DDPO, DDPO shows less mean bias on Ali-CCP and AE-FR than UKD. This may be due to the IPS used in DDPO. EVI achieves the lowest mean bias across the three datasets. Specifically, EVI attains a lower mean bias than EVI w/o VIE, indicating that VIE helps EVI to exploit more unclick-space information.

\subsection{Ablation Study}

The proposed EVI leverages conditional entire-space CVR teacher (CECT) and variational information exploitation (VIE) as its key components. We also perform ablation studies on the CECT and VIE. The ablation study (see Table \ref{Ablation CVR}) shows that EVI performs optimally with both CECT and VIE. In six datasets, including only CECT enhances EVI's performance, but the combination of CECT and VIE achieves the best results. The results highlight the significance of using conditional entire-space CVR teacher and variational information exploitation to improve model performance.

\subsection{Parameter Sensitivity Analysis}
We conduct a sensitivity analysis on EVI to assess the effects of the variational information exploitation loss ratio and the number of transfer layers, as illustrated in Figure \ref{Parameter sensitivity}. For Ali-CCP, AE-FR, AE-ES, and the industrial dataset, performance improves as the variational information loss ratio increases up to 1, whereas AE-NL exhibits an opposite trend. Performance improves with more transfer layers as they convey more useful information.

\section{Conclusion}
This work proposes an entire-space variational information exploitation (EVI) framework to mitigate sample selection bias and data sparsity problems in the CVR estimation task. The EVI ensures unbiased training of the CVR estimator across the entire impression space. To address the bias in pseudo labels of unclicked samples, EVI trains a conditional entire-space CVR teacher model to generate unbiased pseudo labels. Given the limited information carried by pseudo labels, EVI combines logits distillation and variational information lower optimization to enhance the exploitation of non-click space information, thereby improving performance. We conduct extensive experiments on six real-world datasets to validate the effectiveness of our method. In general, EVI shows an average AUC improvement of 2.25\% and a 2.78\% NLL reduction over the optimal baselines. However, EVI relies on accurate CTR estimation and is sensitive to noisy clicked samples. We'll explore methods to mitigate this issue.

\appendix
\section{Experimental Settings}

\subsection{Datasets}
\begin{itemize}
    \item \textbf{Public Dataset.} 
    We validate our method using five public datasets from two e-commerce platforms used in previous related studies: the Alibaba click and conversion prediction dataset (Ali-CCP) \footnote{https://tianchi.aliyun.com/dataset/408}\cite{ma_entire_2018} and the Ali-Express dataset \footnote{https://tianchi.aliyun.com/dataset/74690}\cite{peng2020improving}. The Ali-CCP dataset is obtained from Taobao's online recommender system. The Ali-Express dataset is sourced from the search system of AliExpress, including data from Spain (AE-ES), France (AE-FR), the Netherlands (AE-NL), and the United States (AE-US). Across these five datasets, the user actions follow a sequential pattern of \textit{impression} $\rightarrow$ \textit{click} $\rightarrow$ \textit{conversion}. For Ali-CCP data preprocessing, we use the method from https://github.com/xidongbo/AITM. For Ali-Express, we use the preprocessing method from https://github.com/easezyc/Multitask-Recommendation-Library. Details of these public datasets are presented in Table \ref{dataset table}.
    
    \item \textbf{Industrial Dataset.}
    We validate our method using an industrial dataset from one of the largest affiliate advertising systems in the world. Following principles from previous studies \cite{ma_entire_2018, ma_modeling_2018, wang_escm2_2022}, we use the first four days of data for training and the last day for testing. This approach accounts for daily variability, ensuring a random test set. In industrial datasets, user actions also follow a sequential pattern of \textit{impression} $\rightarrow$ \textit{click} $\rightarrow$ \textit{conversion}. Details of the industrial dataset are presented in Table \ref{dataset table}.
    
\end{itemize}

\subsection{Comparison Methods}
We evaluate our method against 7 representative baseline models in CVR estimation, covering both causal and non-causal approaches. The competitors are as follows:

\begin{itemize}
  \item \textbf{ESMM} \cite{ma_entire_2018}: ESMM applies entire-space modeling to estimate CVR using a CTCVR task.
  
  \item \textbf{ESCM$^2$-IPS} \cite{wang_escm2_2022}: ESCM$^2$-IPS incorporates inverse propensity score (IPS) weighting into ESMM.
  
  \item \textbf{ESCM$^2$-DR} \cite{dai2022generalized}: ESCM$^2$-DR incoperates doubly robust (DR) learning into ESMM.
  
  \item \textbf{MRDR-GPL} \cite{zhou2023generalized}. MRDR-GPL improves the MRDR method \cite{guo_enhanced_2021} by adding generalized propensity learning (GPL) losses, which reduce bias and variance more effectively.
    
  \item \textbf{DR-V2} \cite{li2023propensity}. DR-V2 is an enhanced version of the DR method. It uses balanced-mean-squared-error (BMSE) to improve propensity balancing and reduce variance.
  
  \item \textbf{UKD} \cite{xu2022ukd}. 
  UKD introduces a click adaptive CVR estimator to generate pseudo labels for the entire-space CVR estimator.
  
  \item \textbf{DCMT} \cite{zhu2023dcmt}. DCMT introduces an non-conversion estimator and counterfactual prior knowledge regularizer to build entire-space CVR estimator.
  \item \textbf{DDPO} \cite{su2024ddpo}. DDPO introduces a na\"ive CVR estimator to produce pseudo labels for the entire-space CVR estimator.
\end{itemize}

\subsection{Implementation Details}

All multi-task CVR estimators have similar structures: an embedding layer, a feature extractor, and predictors. The embedding size for all datasets is set to 5. EVI uses MMOE as its feature extractor, while other methods follow their original paper implementations. We use multilayer perceptrons (MLP) with hidden layer sizes of [128, 64, 32] for representation learner. The dimension and number of experts is [256] and 8. The batch size is 8000. We optimize the models using the Adam optimizer \cite{kingma2014adam} with a learning rate of $1e^{-3}$, a weight decay of $1e^{-6}$, $\beta_1=0.9$, and $\beta_2=0.999$. 
We randomly search for all trade-offs in the interval [0, 2] with steps of 0.2 and select the trade-offs that result in the optimal average AUC over five rounds. For the Ali-CCP dataset, we set the parameters as follows: $\lambda_c = 1$, $\lambda_t = 0.2$, $\lambda_r = 2$, $\lambda_i = 0.2$, and $\lambda_g = 0.1$. For other datasets, we set the parameters as $\lambda_c = 0.2$, $\lambda_t = 0.2$, $\lambda_r = 2$, $\lambda_i = 0.2$, and $\lambda_g = 0.2$. The transfer layer number is set to 3. We train all models on Nvidia GeForce RTX 2080Ti GPUs with PyTorch 2.0.

\begin{table}[htbp]
\centering
\caption{Experimental datasets}
\begin{tabular}{ccccc} 
\toprule
Dataset                     & Split & \#Impression & \#Click & \#Conversion  \\ 
\hline
\multirow{2}{*}{Ali-CCP}    & Train & 42.3M      & 1.6M    & 9K            \\
                            & Test  & 43M        & 1.7M    & 9.4K          \\
\multirow{2}{*}{AE-FR}      & Train & 18.2M      & 0.34M   & 9K            \\
                            & Test  & 8.8M       & 0.2M    & 5.3K          \\
\multirow{2}{*}{AE-ES}      & Train & 22.3M      & 0.57M   & 12.9K         \\
                            & Test  & 9.3M       & 0.27M   & 6.1K          \\
\multirow{2}{*}{AE-US}      & Train & 20M        & 0.29M   & 7K            \\
                            & Test  & 7.5M       & 0.16M   & 3.9K          \\
\multirow{2}{*}{AE-NL}      & Train & 12.2M      & 0.25M   & 8.9K          \\
                            & Test  & 5.6M       & 0.14M   & 4.9K          \\
\multirow{2}{*}{Industrial} & Train & 11.78M     & 3.03M   & 0.157M        \\
                            & Test  & 3.31M      & 0.663M  & 55.6K         \\
\bottomrule
\end{tabular}

\label{dataset table}
\end{table}
\section{Proof}
\subsection{Proof of Theorem 1}
\renewcommand{\thetheorem}{1}\begin{theorem}
    The entire-space CVR estimator is biased when the pseudo conversion labels of unclicked samples $r^*_{u,i}$ are biased, i.e., 
\begin{equation}
\left|\mathbb{E}_{\mathcal{D}}\left\lbrack\mathcal{L}_{\mathrm{CVR}}^{\mathrm{entire}}\right\rbrack-\mathcal{L}_{\mathrm{CVR}}^{ideal}\right|>0. \notag
\end{equation}
\end{theorem}

\begin{proof}
\begin{equation}
\begin{aligned} & \left|\mathbb{E}_{\mathcal{D}}\left\lbrack\mathcal{L}_{\mathrm{CVR}}^{\mathrm{entire}}\right\rbrack-\mathcal{L}_{\mathrm{CVR}}^{ideal}\right| & \\  & =\left|\frac{1}{|\mathcal{D}|}\sum_{(u,i)\in\mathcal{D}}\mathbb{E}_{_{\mathcal{D}}}[\delta(r_{u,i}^{},\hat{r}_{u,i})]-\frac{1}{|\mathcal{D}|}\sum_{(u,i)\in\mathcal{D}}\delta(r_{u,i},\hat{r}_{u,i})\right| & \\  & =\left|\frac{1}{|\mathcal{D}|}\sum_{(u,i)\in\mathcal{O}}\delta(r_{u,i},\hat{r}_{u,i})+\frac{1}{|\mathcal{D}|}\sum_{(u,i)\in\mathcal{N}}\delta(r_{u,i}^{*},\hat{r}_{u,i})\right. & \\  & \left.-\frac{1}{|\mathcal{D}|}\sum_{(u,i)\in\mathcal{D}}\delta\left(r_{u,i},\hat{r}_{u,i}\right)\right| & \\  & \stackrel{}{=}\left|\frac{1}{|\mathcal{D}|}\sum_{(u,i)\in\mathcal{N}}\left(\delta\left(r_{u,i}^{*},\hat{r}_{u,i}\right)\ -\delta\left(r_{u,i},\hat{r}_{u,i}\right)\right)\right|\stackrel{(1)}{>}0. & \\  &  & \end{aligned} \notag
\end{equation}
where (1) holds when $r^*_{u,i}$ are biased, i.e., $r^*_{u,i}-r_{u,i}>0$.
\end{proof}

\subsection{Proof of Theorem 2}

\renewcommand{\thetheorem}{2}\begin{theorem}
The conditional entire-space CVR teachers are unbiased, provided that $\hat{o}_{u,i}$ is accurate, i.e.,
\begin{equation}
\mathbb{E}_{\mathcal{D}}[\mathbb{P}(r_{u,i}=1| \hat{O}_{u,i}=\hat{o}_{u,i}] = \mathbb{E}_{\mathcal{D}}[\mathbb{P}(r_{u,i}=1|o_{u,i}=1)]. \notag
\end{equation}
\end{theorem}
\begin{proof}
Given that the CTR distribution is accurately captured by the model, pCTR $\hat{o}_{u,i}$ can be regarded as a propensity score.  Based on Theorem 3 in \cite{rosenbaum1983central}, when conditioned on the propensity score, the observed conversion feedback $r_{u,i}(1)$ and the unobserved conversion feedback $r_{u,i}(0)$ are independent of the click, that is:

\begin{equation}
        \{r_{u,i}(1), r_{u,i}(0)\} \perp  \!\!\!\perp o_{u,i} \mid \hat{o}_{u,i} \notag
\end{equation}

The conditional entire-space CVR teacher uses $\hat{o}_{u,i}$ as a constant feature input to condition the propensity score. This approach is equivalent to making the unobserved conversion feedback $r_{u,i}(0)$ missing at random, thus ensuring the unbiasedness of CVR estimation.

\begin{equation}
    \mathbb{E}_{\mathcal{D}}[\mathbb{P}(r_{u,i}=1|\hat{O}_{u,i}=\hat{o}_{u,i}] = \mathbb{E}_{\mathcal{D}}[\mathbb{P}(r_{u,i}=1|o_{u,i}=1)] \notag
\end{equation}

\end{proof}

\subsection{Proof of Theorem 3}
\renewcommand{\thetheorem}{3}\begin{theorem}
    The EVI CVR estimator is unbiased when the click propensity $\hat{o}_{u,i}$ and pseudo conversion labels of unclicked samples $r^*_{u,i}$ are accurate, i.e., \begin{equation}\left|\mathbb{E}_{\mathcal{D}}\left\lbrack\mathcal{L}_{\mathrm{CVR}}^{\mathrm{EVI}}\right\rbrack-\mathcal{L}_{\mathrm{CVR}}^{ideal}\right|=0.\notag
    \end{equation}
\end{theorem}

\begin{proof}
\begin{equation}
\begin{aligned} & \left|\mathbb{E}_{\mathcal{D}}\left[\mathcal{L}_{\mathrm{CVR}}^{\mathrm{EVI}}\right]-\mathcal{L}_{\mathrm{CVR}}^{ideal}\right|\\  & =\left|\frac{1}{2|\mathcal{D}|}\sum_{(u,i)\in\mathcal{D}}\mathbb{E}_{\mathcal{D}}[\frac{o_{u,i}\delta(r_{u,i},\hat{r}_{u,i})}{\hat{o}_{u,i}}]\right.\\  & +\frac{1}{2|\mathcal{D}|}\sum_{(u,i)\in\mathcal{D}}\mathbb{E}_{\mathcal{D}}[\frac{(1-o_{u,i})\delta(r_{u,i}^{*},\hat{r}_{u,i})}{1-\hat{o}_{u,i}}]\\  & \left.-\frac{1}{|\mathcal{D}|}\sum_{(u,i)\in\mathcal{D}}\delta\left(r_{u,i},\hat{r}_{u,i}\right)\right|\\  & \overset{(1)}{\operatorname*{\operatorname*{=}}}\left|\frac{1}{|\mathcal{D}|}\sum_{(u,i)\in\mathcal{D}}\left(\delta(r_{u,i},\hat{r}_{u,i})-\delta\left(r_{u,i},\hat{r}_{u,i})\right)\right.\right|=0,\end{aligned}\notag
\end{equation}
where (1) holds given that the click propensity $\hat{o}_{u,i}$ and the pseudo conversion labels of unclicked samples $r^*_{u,i}$ are accurate (Theorem 2), i.e.,
$$\mathbb{E}_{\mathcal{D}}[o_{u,i}] = \mathbb{E}_{\mathcal{D}}[\hat{o}_{u,i}],$$
$$r^*_{u,i} = r_{u,i}.$$

\end{proof}

\section*{Acknowledgements}
This work was supported in part by the National Natural Science Foundation of China under Grant 62176042, in part by TCL Technology Innovation Funding SS2024105, and in part by the Fundamental Research Funds for the Central Universities (UESTC) under Grant ZYGX2024Z008.

\bibliography{aaai25}

\end{document}